\documentclass[%
 reprint,
 amsmath,amssymb,
 aps,
 pre,
]{revtex4-1}

\usepackage{graphicx}
\usepackage{dcolumn}
\usepackage{bm}
\usepackage{color} 
\usepackage{ulem} 
\renewcommand{\emph}[1]{{\it #1}}
\usepackage[colorlinks=true,allcolors=blue]{hyperref}
\usepackage{multirow}

\renewcommand{\d}{\mathrm{d}}
\renewcommand{\vec}[1]{\bm{#1}}
\newcommand{\aref}[1]{Appendix~\ref{#1}}
\newcommand{\asref}[1]{Appendices~\ref{#1}}
\newcommand{\sref}[1]{Section~\ref{#1}}
\newcommand{\ssref}[1]{Sections~\ref{#1}}
\newcommand{\eref}[1]{Eq.~\eqref{#1}}
\newcommand{\esref}[1]{Eqs.~\eqref{#1}}
\newcommand{\seref}[1]{\eqref{#1}}
\newcommand{\esrref}[2]{Eqs.~\eqref{#1}--\eqref{#2}}
\newcommand{\fref}[1]{Fig.~\ref{#1}}
\newcommand{\fsref}[1]{Figs.~\ref{#1}}
\newcommand{\sfref}[1]{\ref{#1}}
\newcommand{\tref}[1]{Table~\ref{#1}}

\begin{document}

\title{A geometrically controlled rigidity transition in a model for confluent 3D tissues}

\author{Matthias Merkel}
\affiliation{Department of Physics, Syracuse University, Syracuse, NY 13244, USA}
\author{M.\ Lisa Manning}
\email{mmanning@syr.edu}
\affiliation{Department of Physics, Syracuse University, Syracuse, NY 13244, USA}
\date{\today}

\begin{abstract}
The origin of rigidity in disordered materials is an outstanding open problem in statistical physics. Previously, a class of 2D cellular models has been shown to undergo a rigidity transition controlled by a mechanical parameter that specifies cell shapes. Here, we generalize this model to 3D and find a rigidity transition that is similarly controlled by the preferred surface area: the model is solid-like below a dimensionless surface area of $s_0^\ast\approx5.413$, and fluid-like above this value. We demonstrate that, unlike jamming in soft spheres, residual stresses are necessary to create rigidity. 
These stresses occur precisely when cells are unable to obtain their desired geometry, and we conjecture that there is a well-defined minimal surface area possible for disordered cellular structures. 
We show that the behavior of this minimal surface induces a linear scaling of the shear modulus with the control parameter at the transition point, which is different from the scaling observed in particulate matter.
The existence of such a minimal surface may be relevant for biological tissues and foams, and helps explain why cell shapes are a good structural order parameter for rigidity transitions in biological tissues.
\end{abstract}

\maketitle 

\section{Introduction}
Many biological tissues are confluent, where there are no gaps or overlaps between cells. These tissues are active, disordered, far-from equilibrium materials, sharing similarities with both fiber networks and disordered particulate or glassy matter. Recent experiments have demonstrated that confluent tissues can exhibit a glassy fluid-to-solid transition~\cite{Angelini2011, Sadati2013, Schoetz2013,  Nnetu2012}, which likely plays a role in development and disease~\cite{Park2015,Pawlizak2015}. Therefore, an interesting open question is how a cell's structure and mechanics influence the rigidity of a confluent tissue.

The onset of rigidity in single cells~\cite{Ingber1997} and groups of cells below confluence~\cite{Belmonte2008,Henkes2011} has been fruitfully studied using simple tensegrity and particle models. To understand rigidity in confluent tissues, we study a simple vertex model~\cite{Honda1984,Farhadifar2007,Fletcher2014,Bi2014,Bi2015,Su2016,Barton2016,Alt2017} that incorporates the constraints on cell shape imposed by confluence.  The original models focused on 2D monolayers of cells, which are described as networks of cellular polygons that tessellate space, where the degrees of freedom are the vertices of the polygons. Cells pay an energetic penalty when their perimeter differs from a preferred value $P_0$ and their cross-sectional area differs from a preferred area $A_0$. Variations include replacing edges and vertices by a Voronoi tessellation of cell centers~\cite{Bi2016,Su2016,Barton2016,Sussman2017}, and including active dynamics \cite{Sato2015,Bi2016,Barton2016}. Although the dynamical rules for the evolution of cellular potts models are quite different, most also use a similar energy functional based on interfacial energies~\cite{Graner1992,Kabla2012}.

Recently, it was discovered that such models exhibit rigidity or glass-like transitions controlled by the non-dimensionalized preferred perimeter $p_0=P_0/\sqrt{A_0}$~\cite{Farhadifar2007,Staple2010,Bi2014,Bi2015,Bi2016}. While the hexagonal ground state becomes linearly unstable if $p_0>3.72$~\cite{Farhadifar2007,Staple2010}, there is a bona fide second order rigidity transition in disordered structures with a critical point reported at $p_0^\ast=3.81$~\cite{Bi2014,Bi2015}. Adding active self-propulsion converts this point into a line of glass-like transitions in vertex models~\cite{Bi2016} and cellular potts models~\cite{Chiang2016}. A priori model predictions for cell shapes have recently been verified in experiments on glassy 2D monolayers~\cite{Park2015}. As in many other systems, the critical point controls global mechanical properties and collective motion, which both contribute to biological function.

But what is the origin of this transition, and why is it so robust -- occurring in both simulations (vertex, Voronoi, cellular potts) and experiments? In both particulate granular matter and fiber networks, rigidity  is fairly well understood~\cite{Sarkar2013,Kamien2007}. For example, frictionless athermal particles rigidify or jam at a critical value of the packing fraction, and the onset of rigidity is predicted by Maxwell's constraint counting rule \cite{Maxwell1864,Calladine1978,Liu2010}. In models for fiber networks, the bond occupation probability controls the transition point also according to Maxwell's criterion~\cite{Jacobs1995,Ellenbroek2015,Lubensky2015}.  Moreover, exploring the effect of dimensionality~\cite{Charbonneau2017} has helped to discriminate between different theories for the origin of rigidity. Therefore it is natural to study vertex models in different dimensions to investigate whether constraint counting explains rigidity in confluent tissues. 

\begin{figure*}
  \includegraphics{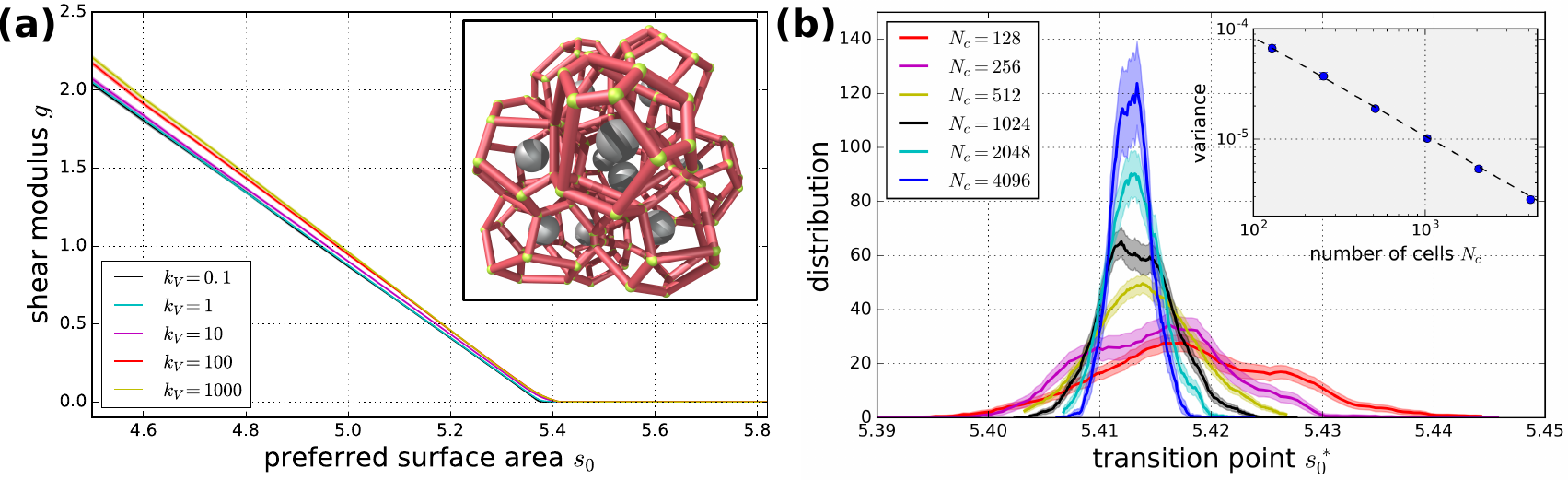}
  \caption{
  (a) Existence of a rigidity transition.  The average shear modulus $g$ vanishes when the preferred surface area $s_0$ is approximately $5.41$, and changes only slightly as a function of the volume rigidity $k_V$.
  (a,~inset) Cell shapes are defined by the Voronoi tessellation of the cell positions $\vec{R}_i$ (gray spheres).
  (b) Finite size scaling for the transition. Smoothed distribution of the finite-size transition points $s_0^\ast$, determined from the fraction $F(s_0)$ of rigid energy-minimized configurations with $k_V = 10$. Details of the smoothing and identification of the transition point are given in \asref{sec:transitionPoint} and \ref{sec:fss}. The shaded regions indicate the standard error of the mean.
  (b,~inset) Variance of the distribution $p$ of transition points, computed directly from the fractions of rigid networks $F(s_0)$.  Error bars indicating the standard error of the mean are within symbol size.  The dashed line shows a power law fit with exponent $-0.90\pm0.04$.
  \label{fig:modelAndTransition}}
\end{figure*}
Furthermore, although curved 2D cell sheets embedded in 3D space~\cite{Honda2008a,Hannezo2014,Monier2015,Okuda2015,Bielmeier2016,Misra2016} are an an active area of research, there is surprisingly little work on confluent bulk 3D tissues \cite{Honda2004,Viens2007}, and that work predates the discovery of the rigidity transition in 2D. While there are currently few experimental observations of glassy behavior in fully 3D tissues~\cite{Schoetz2013}, rapid advances in microscopy and segmentation algorithms~\cite{Keller2013,Stegmaier2016} are currently generating a host of data on 3D bulk tissues that could be used to test model predictions and make connections to embryonic development and disease {\it in vivo}.

Here we generalize an existing 2D Voronoi model for isotropic tissues~\cite{Bi2016} to three dimensions, and find a rigidity transition controlled by the dimensionless preferred shape index with an associated structural order parameter. We demonstrate that in contrast to jammed solids, residual stresses that arise when a cell is unable to attain its preferred geometry rigidify the system. Because the mechanisms for rigidity in vertex and particle-based models are distinct, we can identify several specific differences in response and structure that could be used to distinguish between models in real tissues.

In addition, finite-size scaling indicates that the transition occurs at a precise value of the 3D shape index, leading us to conjecture that there is a well-defined minimal surface for cellular structures that controls this rigidity transition. Our work suggests that the origin of rigidity is purely geometric and does not depend on the details of the vertex energy functional. This provides a possible explanation for why observations of rigidity transitions are so robust in cellular systems, and it strongly suggests that our results hold not only for our specific model but for a much broader class of 3D tissue models.

\section{Model and Methods}
We describe a three-dimensional confluent tissue by a network of $N_c$ cells, where each cell $i$ is represented by a position vector $\vec{R}_i$ (\fref{fig:modelAndTransition}a, inset).  Cell shapes are described by the Voronoi tessellation of these cell positions. In a straightforward generalization of 2D Voronoi models~\cite{Bi2016} to 3D, inter-cellular forces are defined as derivatives on an effective energy functional:
\begin{equation}
	E = \sum_i{\bigg[K_V(V_i-V_0)^2+K_S(S_i-S_0)^2\bigg]}\text{,}\label{eq:energy}
\end{equation}
where $V_i$ and $S_i$ correspond to cell volumes and surface areas, respectively. The sum is over all cells of the network, $K_V>0$ is the volume rigidity, $V_0$ is the preferred volume, $K_S>0$ is the surface rigidity, and $S_0$ is the preferred surface area. 

Our goal is to simulate this model to determine if there is in fact a rigidity transition in 3D, and to shed light on the origin of rigidity.  This is a technically challenging question, as we need to identify local energy minima, which correspond to states in mechanical equilibrium, sampling over instantiations of the disorder. We also need to determine precisely if the shear modulus at the minima are distinct from zero over a wide range of model parameters. To perform these tasks, we need very accurate numerical calculations of the (rather complicated) first and second derivatives of the energy functional given by ~\eqref{eq:energy}. Therefore, in \aref{sec:analytic}, we present details required to develop analytic expressions for these derivatives, which serves two purposes. First, it allows the most accurate computations, and second, it allows us to compare the analytic and numerical derivatives as a consistency check to validate our code.

\begin{figure*}
  \includegraphics{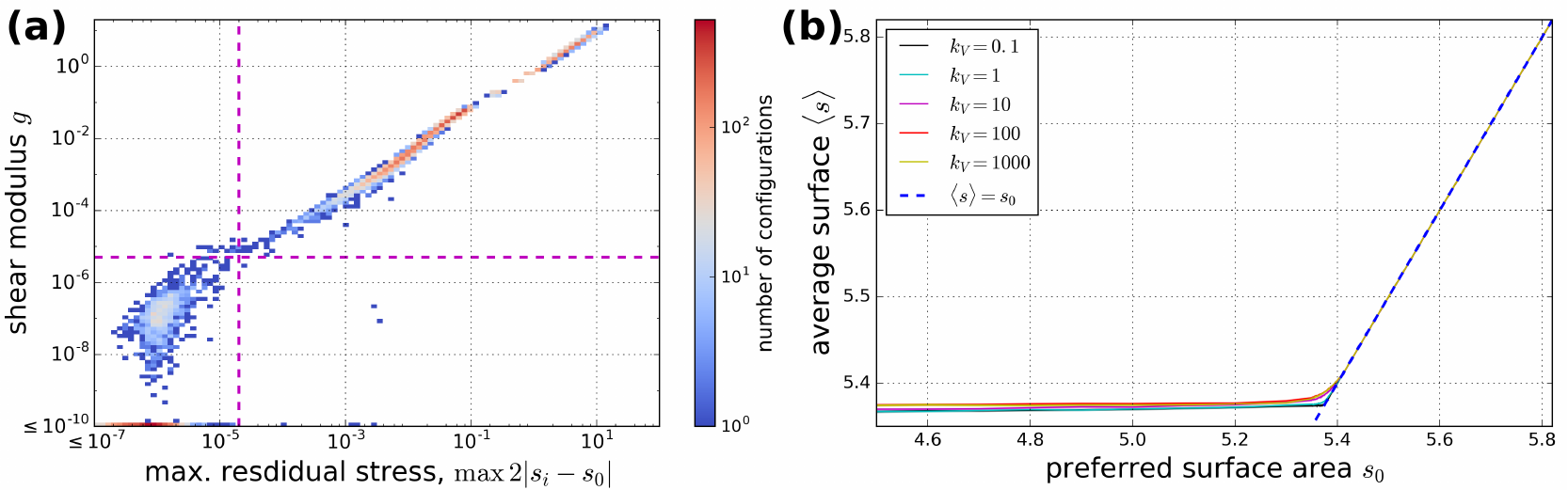}
  \caption{(a) Residual stresses are both necessary and sufficient for rigidity.   Histogram characterizing all energy-minimized states for $k_V=1$ and all values of $s_0$ as a function of shear modulus $g$ the maximal surface tension magnitude $2\lvert s_i-s_0\rvert$.  The magenta dashed lines indicate cutoffs on shear modulus and max.\ surface tension.  See \aref{sec:residualStressesCreateRigidity} for the corresponding plot with the pressure on the $x$ axis.
  (b)  Average cell surface area $\langle s\rangle$ as a function of the preferred surface area $s_0$ for different volume rigidities $k_V$. All surface areas and volumes match their preferred values above the transition point $s_0>s_0^\ast$.  
  \label{fig:residualstressesAndMinSurface}}
\end{figure*}
We use periodic boundary conditions with a fixed cubic box size of $L^3$.  As in 2D, the preferred volume $V_0$ only renormalizes the pressure and does not affect the forces between the cells (\aref{sec:totalVolume}) \cite{Su2016,Yang2017}.
Therefore, without loss of generality we set $V_0 = \langle V\rangle = L^3/N_c$ and non-dimensionalize our model with respect to the length unit $\langle V\rangle^{1/3}$ and the energy unit $K_S\langle V\rangle^{4/3}$, leading to the dimensionless energy:
\begin{equation}
	e = \sum_i{\bigg[k_V(v_i-1)^2+(s_i-s_0)^2\bigg]}\text{.}\label{eq:energy-dimensionless}
\end{equation}
The remaining three dimensionless parameters are the preferred shape index $s_0=S_0/\langle V\rangle^{2/3}$, the relative volume rigidity $k_V=K_V\langle V\rangle^{2/3}/K_S$, and the number of cells $N_c$. 

Extended details of the initialization and minimization are described in \aref{sec:implementation}. Briefly, unless otherwise noted, we fix $N_c=512$. We vary $k_V$ logarithmically and $s_0$ linearly between $0$ and the ideal gas value $5.82$, which is the average cell surface area $\langle s\rangle=(\sum_i{s_i})/N_c$ for randomly placed cell centers~\cite{Meijering1953} (\aref{sec:parameters}).  Reported values are averaged over 100 random initial configurations minimized to a local energy minimum using the BFGS algorithm~\cite{Fletcher2000}, varying the cell positions $\vec{r}_i$ and the simple shear degree of freedom $\gamma$ (\aref{sec:energyMinimization}).

\section{Results}
\subsection{Existence of the rigidity transition}
For each relaxed state, we calculate the simple shear modulus $g=(\d^2 e/\d\gamma^2)/N_c$ using an analytic expression based on the Hessian matrix that describes the second derivatives of the energy functional (\asref{sec:hessianMatrix}--\ref{sec:shearModulus}). We find that there is a continuous transition in $g$ at a preferred relative surface area of $s_0=s_0^\ast\approx 5.4$ with a solid regime ($g>0$) for $s_0$ below this transition point and a fluid regime ($g=0$) above it (\fref{fig:modelAndTransition}a).  Neither the transition point $s_0^\ast$ nor shear modulus $g$ depend strongly on $k_V$. 
We also studied the precise distribution of transition points $p(s_0^\ast)$ (\fref{fig:modelAndTransition}b) (\aref{sec:transitionPoint}).
We found a finite variance of this distribution, which appears to be only due to finite-size effects (\fref{fig:modelAndTransition}b inset) and identify an extrapolated transition point of $s_0^\ast\approx5.413$ in the limit $N_c\rightarrow\infty$ (\aref{sec:fss}).

\subsection{Mechanism that generates rigidity}
To better understand the origin of the rigidity transition, we begin with simple constraint counting which has successfully been used to predict rigidity in many systems, including spring networks~\cite{Lubensky2015,Vermeulen2017}.  Our system has $3N_c+1$ degrees of freedom, 3 for each cell position plus the shear degree of freedom. If each of the $2N_c$ terms in \eref{eq:energy-dimensionless} is regarded as a \emph{generalized} spring and thus contributes a single constraint, the system would be highly under-constrained with at least $N+1$ zero modes for all values of $s_0$, inconsistent with our observations.

To understand how this seemingly under-constrained system can rigidify, we define the extended Hessian $\bar{D}_{pq}$:
\begin{equation}
\begin{aligned}
  \bar{D}_{pq} &= 2\sum_i{}\Bigg[\frac{\partial s_i}{\partial z_p}\frac{\partial s_i}{\partial z_q} + k_V\frac{\partial v_i}{\partial z_p}\frac{\partial v_i}{\partial z_q} \\
  &\qquad\; + (s_i-s_0)\frac{\partial^2 s_i}{\partial z_p\partial z_q} + k_V(v_i-1)\frac{\partial^2 v_i}{\partial z_p\partial z_q}\Bigg]\text{,}
\end{aligned}\label{eq:dynamicalMatrix}
\end{equation}
where $(z_p)$ is a $(3N_c+1)$-dimensional vector of the cell positions and the shear degree of freedom, and the sum is over all cells $i$. The full analytic expression of the extended Hessian is derived in \aref{sec:hessianMatrix}. 
The constraints imposed by the $2N_c$ generalized springs discussed above correspond to the first two terms in \eref{eq:dynamicalMatrix}, while the last two terms correspond to residual stresses, which are known to rigidify otherwise under-constrained systems \cite{Calladine1978,Ingber1997,Alexander1998,Vermeulen2017}. In our system, residual stresses are the $2 N_c$ surface tensions $2(s_i-s_0)$ and pressures $2k_V(1-v_i)$. 

To study whether residual stresses are both necessary and sufficient to generate rigidity,  we plot a two-dimensional histogram categorizing energy-minimized states with respect to both their shear modulus $g$ and the maximal surface tension magnitude of all cells within the configuration, \fref{fig:residualstressesAndMinSurface}a. A similar histogram also holds for the pressure (\fref{fig:g-residualStresses}a, \aref{sec:residualStressesCreateRigidity}).
The fact that the upper left quadrant is largely devoid of configurations shows that residual stresses are necessary for rigidity. The few exceptions are close to both cutoff values and likely due to imperfect minimization.

We independently verified this by computing the overlap between the shear degree of freedom and the infinitesimal zero modes, which is always finite (\aref{sec:overlap}).  This shows that the system would indeed have no resistance to shear in the absence of residual stresses. 

The lower right quadrant in \fref{fig:residualstressesAndMinSurface}a is also devoid of configurations, except for a handful of systems (about 1 in 1000). These are floppy except for a single rigid cell of a specific polyhedron type with anomalously few neighbors  -- a 10-sided truncated square trapezohedron (\fref{fig:g-residualStresses}b). Hence, up to these very few exceptions, the rigidity transition is created by the onset of residual stresses, which typically occurs in every cell simultaneously (\aref{sec:collectiveOnset}).

So what controls the onset of residual stresses in our system?
In the floppy regime, when all surface tensions $2(s_i-s_0)$ and pressures $2k_V(1-v_i)$ are zero,  cells exactly attain their desired shapes so that the average observed shape index $\langle s\rangle$ equals the preferred shape index $s_0$, as shown in  \fref{fig:residualstressesAndMinSurface}b and there is zero standard deviation in cell surfaces and volumes ($\sigma_s=\sigma_v=0$ with $\sigma_s^2=(\sum_i{[s_i-\langle s\rangle]^2})/N_c$ and $\sigma_v$ defined analogously, see \fref{fig:sigmas}). We call this geometric compatibility. 
Because the energy functional \eref{eq:energy-dimensionless} drives residual stresses towards zero, the fact that residual stresses are non-zero in the solid indicates that no geometrically compatible state is reachable by standard energy minimization. This in turn suggests that the transition point $s_0^\ast$ is determined by a purely geometric criterion: it corresponds to a local minimum in the average surface area $\langle s\rangle$ under the constraint that there are no surface and volume fluctuations $\sigma_s=\sigma_v=0$ as in the fluid. 

Thus, based on our finite-size scaling analysis (\fref{fig:modelAndTransition}b), we conjecture that in the thermodynamic limit there is a minimum possible value of the average surface area $\langle s\rangle$ for disordered Voronoi tessellations with $\sigma_s=\sigma_v=0$, which is given by $s_0^\ast\approx5.413$.  

This conjecture is reminiscent of those for jammed packings of particles, where the distribution of jamming packing fractions approaches a narrowly peaked function in the limit of large system sizes~\cite{Xu2005,Chaudhuri2010, Charbonneau2012}.
It has been suggested that this packing fraction (about 64\% in 3D) can be defined as the maximally disordered rigid state of spheres~\cite{Torquato2000}, or a divergence in the rate at which accessible disordered states disappear~\cite{Kamien2007}. 

As in theories for jamming, we must specify that we are restricting ourselves to random tessellations, as $\langle s\rangle$ can become smaller than $s_0^\ast$ for ordered states:  the Voronoi packing corresponding to the ordered Kelvin structure has $\langle s\rangle \approx 5.315$ (\aref{sec:periodicPackings}).
Identifying which configurations lie on the ``disordered'' branch is of course difficult. In jamming, it is clear that different protocols for generating packings on the ``disordered'' branch generate distinct critical packing fractions, although the numbers are all still remarkably close~\cite{Chaudhuri2010}; and we find something similar here. For example, when using a conjugated gradient minimizer instead of BFGS, we have found a disordered transition point slightly shifted down by $\lesssim 0.01$.  A weak dependence on the minimization protocol was recently also observed for the 2D Voronoi model~\cite{Sussman2017}.

\begin{figure*}
  \centering
  \includegraphics{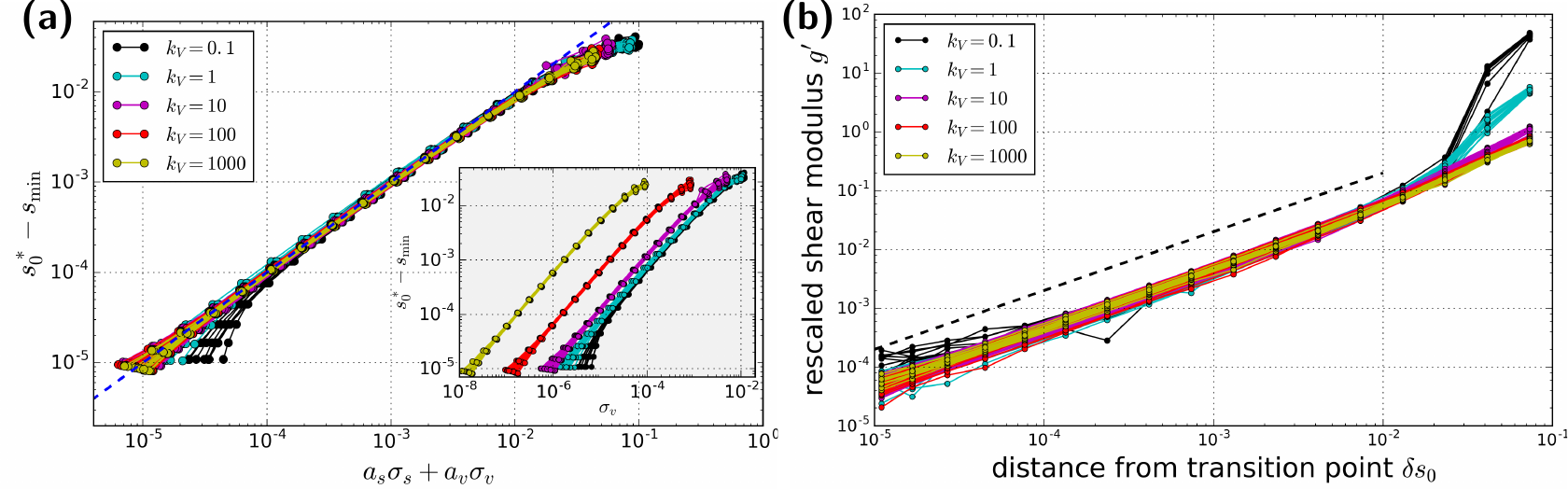}
  \caption{
  Scaling behavior of the model in the solid vicinity of the transition point.
  (a) The average surface area $\langle s\rangle=s_\mathrm{min}$ is reduced below the transition value $s_0^\ast$ as fluctuations in surface area $\sigma_s$ and volume $\sigma_v$ emerge.  
  We can collapse all the data for different $k_V$ between $0.1$ and $1000$ onto the line $s_0^\ast - \langle s\rangle = a_s\sigma_s + a_v\sigma_v$ with $a_s=2.0$ and $a_v=6.8$ (blue dashed line) \cite{MerkelPRE}. 
  (a,~inset) Plotting $s_0^\ast - \langle s\rangle$ over the volume standard deviation $\sigma_v$ shows no collapse.
  (b) The shear modulus scales linearly with the distance from the transition point $\delta s_0=s_0^\ast-s_0$.  In particular, the shear modulus rescaled by a constant, $g'=(1+a_s^2+a_v^2/k_V)g$, shows a collapse for the different values of $k_V$ when plotted with respect to $\delta s_0$.
  The black dashed line indicates linear scaling.
  For both panels, we used a set of dedicated simulations that explore the solid vicinity of the transition point for a given realization (\aref{sec:solidVicinity}).  Different curves of the same color correspond to different initial conditions.  Deviations from linear behavior appearing in both panels close to the transition point for small $k_V$ are related to the finite shear modulus cutoff involved in determining the transition point $s_0^\ast$ for a given configuration.
  \label{fig:linearScaling}}
\end{figure*}
\subsection{Universal behavior and scaling of the shear modulus}
\label{sec:scaling}
\fref{fig:residualstressesAndMinSurface}b also shows that in the solid regime, the average surface area $\langle s\rangle$ decreases below the transition point $s_0^\ast\approx5.413$, as the surface and volume fluctuations rise away from zero (\fref{fig:sigmas}).   
This is not surprising, as \eref{eq:energy-dimensionless} can be rewritten:
\begin{equation}
  e = N_c\Bigl[\big(\langle s\rangle - s_0\big)^2 + \sigma_s^2 + k_V\sigma_v^2\Bigr]\text{.}\label{eq:energy-rewritten}
\end{equation}
In the floppy regime all three terms can be zero simultaneously because geometrically compatible states are attainable, but in the solid this is not possible and we find that the minimal surface is a function of the variances: $\langle s\rangle=s_\mathrm{min}(\sigma_s,\sigma_v)$ \cite{MerkelPRE}.  
In the vicinity of the transition point, $s_\mathrm{min}(\sigma_s,\sigma_v)$ is surprisingly simple and universal: we find numerically that \begin{equation}
  s_\mathrm{min}(\sigma_s,\sigma_v) = s_0^\ast - a_s\sigma_s - a_v\sigma_v\label{eq:sMin}
\end{equation}
with $a_s\approx2.0$ and $a_v\approx6.8$, as shown by the data collapse in \fref{fig:linearScaling}a. This result is independent of the parameters of the energy functional $s_0$ and $k_V$.  Strikingly, the collapse also shows that the parameters $a_s$ and $a_v$ are largely independent of the random initial conditions, suggesting that they are universal geometrical properties of disordered 3D Voronoi packings.  Possibly, a relation like \eref{eq:sMin} with universal coefficients may also hold for general 3D cellular packings. 

This linear scaling in the minimal surface $s_\mathrm{min}$ explains the behavior of the shear modulus close to the transition point, which scales linearly with the distance from the transition point, $\delta s_0 = s_0^\ast-s_0$ (\fref{fig:linearScaling}b).
To understand this, we start from the formula \eref{eq:extHessianShearModulus} for the shear modulus $g$ (\aref{sec:extendedHessian}), which we restate here:  If and only if the extended Hessian $\bar{D}_{pq}$ has a zero mode with a nonzero shear component, then the shear modulus $g$ is zero. Otherwise, the shear modulus can be computed as:
\begin{equation}
  g = \frac{1}{N_c}\left[\sum_{m}{\frac{(\bar{u}_\gamma^m)^2}{\bar{\omega}_m^2}}\right]^{-1}\text{.}\label{eq:computationG}
\end{equation}
Here the sum is over all strictly positive eigenvalues $\bar{\omega}_m^2$ of the extended Hessian $\bar{D}_{pq}$, and $\bar{u}_\gamma^m$ are the shear components of the associated eigenvectors. 
First, we demonstrate that this expression gives the correct answer in the fluid regime, where there are no residual stresses, and the shear modulus should be zero.
Without residual stresses 
there are at least $(N_c+1)$ eigenmodes of $\bar{D}_{pq}$ with eigenvalue zero, corresponding to the $(N_c+1)$ zero modes obtained by the naive constraint counting. As shown in \aref{sec:overlap}, at least one of these zero modes has a nonzero shear component, which is why the system is floppy with $g=0$.

As discussed above, in the solid vicinity of the transition point, the cellular surface tensions $2(s_i-s_0)$ and pressures $2k_V(1-v_i)$ appear and rigidify the system. 
In particular, while the average pressure is zero because $\langle v\rangle=1$, \esref{eq:energy-rewritten} and \seref{eq:sMin} imply that the average surface tension scales linearly with the distance to the transition point: $2(\langle s\rangle-s_0)=2\delta s_0/(1+a_s^2+a_v^2/k_V)$ (\aref{sec:scalingSurfaceTension}).
Moreover, it turns out that the surface tensions and pressures \emph{of each individual cell} are proportional to $\delta s_0$ \cite{MerkelPRE}.  As a consequence, many of the $(N_c+1)$ eigenvalues that were zero in the fluid increase by an amount proportional to $\delta s_0$, and these small eigenvalues then dominate the formula for the shear modulus $g$ according to \eref{eq:computationG}.  Hence, the shear modulus scales linearly with the distance to the transition point $g\sim\delta s_0$. We show this numerically by plotting a rescaled shear modulus $g'=(1+a_s^2+a_v^2/k_V)g$ over $\delta s_0$. The fact that this quantity collapses for different values for $k_V$ demonstrates that the average surface tension $2(\langle s\rangle-s_0)$ is a dominant contribution to shear modulus. This is completely different from what is observed in particulate matter, where constraints are added as particles come into contact, generating a shear modulus that scales as the square root of the dimensionless control parameter (the packing fraction).


\section{Discussion and Conclusions}
To our knowledge, this is one of the first systematic numerical studies of the mechanics of confluent bulk biological tissues \cite{Honda2004,Viens2007}. Our 3D Voronoi-based model, which is a straightforward generalization of successful 2D models for epithelial sheets~\cite{Bi2016}, exhibits a rigidity transition when the dimensionless preferred cell surface area is $s_0^\ast \approx 5.413$. The transition is accompanied by a structural order parameter, the observed average cell surface area $\langle s \rangle$. 

In addition, we finally have a numerics-backed conjecture that explains the origin of rigidity in vertex-like models, as well as an explanation for the robustness of the structural order parameter. In contrast to jamming in frictionless spheres, where an unjammed system acquires more constraints as the packing fraction increases and spheres touch, cells in vertex models always have the same number of surface and volume constraints independent of model parameters.

Instead, rigidity is created by residual stresses which are due to geometric incompatibility. Specifically, we conjecture that there is a minimal surface area possible for disordered cellular structures under the constraint that each cell has an identical surface area and volume.  These constraints arise naturally at the transition point, because in the fluid phase each cell can exactly attain its desired shape and so there are no fluctuations in those quantities.  In the solid, the system would like to have a surface smaller than the minimal one, and so it is stuck there, although small fluctuations in the surface area or volume of each cell can reduce the surface area below the transition value. We believe that this is also the mechanism controlling the transition in 2D vertex models \cite{Bi2016,Sussman2017}.

Although we have focused here on the relevance to biological tissues, it is likely that rigidity transitions driven by residual stresses may appear in a wide variety of other models for disparate physical phenomena~\cite{Vermeulen2017}, and the collective nature of the transition could lead to glassy dynamics that is strikingly different from that seen in particulate matter.

Our finding that a purely geometric quantity governs the onset of rigidity in a vertex model may also help explain why the correlation between cell shape and rigidity is so robust~\cite{Bi2015,Bi2016} and even holds in experiments on biological tissues~\cite{Park2015}. Our results suggest that any system where cellular structures minimize their surface area and suppress fluctuations to their surface areas and volumes could rigidify when their shape index $\langle s\rangle$ drops below $5.413$, since there are no available states below this value.

This leads to an immediate and testable prediction for experiments on 3D cell aggregates including embryonic zebrafish cells~\cite{Schoetz2013} and human breast cancer cells~\cite{Pawlizak2015}, as well as 3D tissue explants from vertebrate embryos~\cite{Hopyan2017}. Our model predicts that in the solid $\langle s \rangle \approx 5.4$, and $\langle s \rangle$ should rise away from that value as tissue fluidizes. 

Although a similar prediction was successful in 2D~\cite{Park2015}, 
additional mechanical interactions not yet included in our model may be more important in 3D than 2D. Therefore, we think of the model presented here as a useful null hypothesis for establishing a relationship between tissue structure and tissue mechanics in 3D; many different perturbations can and should be studied, as they may alter the transition.  For example, we will discuss the influence of cell motility and persistence on the 3D model elsewhere~\cite{MerkelMotility}, but just as in 2D we find that cell shape is still an excellent predictor of tissue rheology even in the presence of motile forces.  Additional useful extensions to the model could account for cell division (which may fluidize the solid phase~\cite{Puliafito2012,Matoz-Fernandez2017}) and polydispersity in the preferred surface areas and volumes. 

Another possible perturbation is nuclear rigidity, as an alternate explanation for tissue rigidification is that the nuclei jam.  Interestingly, the shape index associated with Voronoi cells for particulate matter at jamming is 5.38~\cite{Morse2016}, which very close but distinct from the transition we observe here. 

Importantly, because the nature of the transitions in vertex and particle models are fundamentally different, we can identify several observables that should distinguish between these two mechanisms. In vertex models, fluctuations in Voronoi volumes and surface areas are minimal in the fluid phase and grow in the solid phase, whereas for particulate matter these fluctuations are large in the fluid phase and get smaller as one approaches the solid phase. As discussed in \sref{sec:scaling} we also predict different scaling laws for the shear modulus: in jammed particulate matter the modulus scales with the square root of the control parameter (the packing fraction) \cite{Liu2010b}, while in the vertex model the modulus scales linearly with the control parameter (the target shape index). Finally, there is no bulk modulus in the fluid phase of jammed particles, but always a significant bulk modulus in vertex models, even in the fluid regime.



%

Although we present numerical evidence for the origin of the transition, it would be interesting to try to develop geometric arguments that predict the value of the dimensionless disordered minimal surface $s_0^\ast \approx 5.413$ as well as analytic arguments that explain why the residual stresses are both necessary and sufficient to create rigidity. Recent work by Moshe and collaborators on lattice-based structures develops a nice framework for studying this problem~\cite{Moshe2017}.

It would also be interesting to test the predictions of our model in passive cellular materials like foams or biomimetic cellular materials~\cite{Weaire2001,Kraynik2003,Jorjadze2011}. For example, random foams are deep in the solid phase ($s_0\rightarrow-\infty$ and $k_V/\vert s_0\vert\rightarrow\infty$), and since they seek to minimize their surface area under an evolution of cell volumes driven by mean curvature flow, they should also be governed by our minimal surface hypothesis.  For example, Ref.~\cite{Kraynik2003} reports values for the average shape index $\langle s \rangle$, but not the fluctuations $\sigma_s$. It would be interesting to revisit these results and determine if the surfaces are related to our minimal family $s_\mathrm{min}(\sigma_s,\sigma_v)$.  

Finally, our findings could also be used to create artificial cellular materials that may be hyperuniform~\cite{Bi2016a}, and can transition between solid-like and fluid-like behavior depending on some microscopic parameter, such as an effective surface tension tuned using chemicals, light, or magnetic fields \cite{Brown2013,Jorjadze2011}, or on macroscopic parameters like the overall volume or the pressure, which change the average cell volume.

\begin{acknowledgments}
MM and MLM thank Peter Morse, Daniel M.\ Sussman, and Michael Moshe, as well as Steffen Grosser, J\"urgen Lippoldt, and Josef K\"as for fruitful discussions.  Both authors also acknowledge funding from the Alfred P.\ Sloan Foundation, the Gordon and Betty Moore Foundation, the Research Corporation for Scientific Advancement, and computational support through NSF ACI-1541396. MLM also acknowledges support from the Simons Foundation under grant numbers 446222 and 454947, and NSF-DMR-1352184 and NSF-PHY-1607416. 
\end{acknowledgments}

\appendix
\section{Analytic results that allow efficient computation and analysis of the 3D Voronoi model}
\label{sec:analytic}
\subsection{Effect of preferred volume}
\label{sec:totalVolume}
Here we show that $V_0$ only offsets the pressure but does not affect intercellular forces and the shear modulus.  To this end, we transform \eref{eq:energy} into:
\begin{equation}
\begin{aligned}
  E &= \sum_i{\bigg[K_V(V_i-\langle V\rangle)^2 + K_S(S_i-S_0)^2\bigg]} \\
  	&\qquad\qquad\qquad\qquad\qquad+ \frac{K_V}{N_c}(V_\mathrm{tot}-N_cV_0)^2\text{.}
\end{aligned}
\end{equation}
Here $V_\mathrm{tot}$ is the total volume of the system.  Note that $V_0$ only appears in the last term where it only couples to the total system volume.  As a consequence, $V_0$ does not affect any forces except that it offsets the pressure of the system.  Hence, we just fixed $V_0=\langle V\rangle$ such that the last term disappears.  Note that an analogous argument also holds in 2D \cite{Su2016,Yang2017}.

\subsection{Periodic boundary conditions}
\label{sec:pbc}
To implement the periodic boundary conditions in a clean way, we explicitly expressed the total energy $e$ in terms of distance vectors $\vec{r}_{ij}$ between neighboring cell positions $i$ and $j$ instead of absolute cell positions (see next section, \sref{sec:energy}). 
Separately, these distance vectors $\vec{r}_{ij}$ are expressed in terms of absolute cell positions $\vec{r}_i$ and $\vec{r}_j$. However, the $\vec{r}_{ij}$ also depend on the periodic boundary conditions, because cells on opposing sides of the periodic box are neighbors of each other.
To implement this, each cell neighbor pair $(i,j)$ is given a integer ``periodicity'' vector $\vec{q}_{ij}$.  It is defined such that $\vec{q}_{ij}=0$ whenever one can get from position of cell $i$ to position of cell $j$ without crossing a face of the periodic box.  If one needs to cross for instance the upper/lower face of the periodic box once from below in going from $i$ to $j$, then $q_{ij}^z=+1$, and analogous for the other components of $\vec{q}_{ij}$ (see also the appendix in \cite{Merkel2014b} for an extensive explanation of the analogous 2D case).
The distance vectors $\vec{r}_{ij}=(r_{ij}^x,r_{ij}^y,r_{ij}^z)$ then depend as follows on the cell position and the dimensionless periodic box dimensions $l_x,l_y,l_z$:
\begin{equation}
\begin{aligned}
	r_{ij}^x &= r_j^x - r_i^x + q_{ij}^xl_x \\
	r_{ij}^y &= r_j^y - r_i^y + q_{ij}^yl_y \\
	r_{ij}^z &= r_j^z - r_i^z + q_{ij}^zl_z\text{.}
\end{aligned}
\end{equation}

We are also interested in the effect of simple shear.  We thus allow for ``skewed'' periodic boundary conditions and characterize simple shear by the shear variable $\gamma$.  It is implemented by modifying the above relations as follows:
\begin{equation}
\begin{aligned}
	r_{ij}^x &= r_j^x - r_i^x + q_{ij}^xl_x + \gamma q_{ij}^yl_y \\
	r_{ij}^y &= r_j^y - r_i^y + q_{ij}^yl_y \\
	r_{ij}^z &= r_j^z - r_i^z + q_{ij}^zl_z \text{.}
\end{aligned}\label{eq:skewedPbc}
\end{equation}
Note that this relation between cell distance vectors $\vec{r}_{ij}$ and cell positions $\vec{r}_i$ is the only place where the periodic boundary conditions appear.  This allows to completely separate the physics from the boundary conditions in the implementation.

\subsection{Energy}
\label{sec:energy}
Here, we express the total energy $e$ in terms of the distance vectors $\vec{r}_{ij}$. It is given by the sum of all cell energies $e_i$:
\begin{equation}
	e = \sum_i{e_i},\label{eq:totalEnergy}
\end{equation}
with
\begin{equation}
	e_i = k_V(v_i-1)^2+(s_i-s_0)^2.
\end{equation}
Volume and surface of a cell $i$ are given by:
\begin{align}
  v_i &= \frac{1}{6}\sum_n{\vec{r}_{in}\cdot\vec{a}_{in}}\label{eq:volume}\\
  s_i &= \sum_n{\lvert\bm{a}_{in}\rvert}\label{eq:surface}\text{.}
\end{align}
Here, both sums are over all cells $n$ neighboring cell $i$. The vector $\vec{a}_{in}$ denotes the oriented area of the polygonal interface between cells $i$ and $n$ pointing orthogonally towards $n$. The dot in the volume formula denotes the scalar product and the vertical bars in the surface formula denote the norm of the vector.

We thus need to express the oriented area $\vec{a}_{in}$ of the face between cells $i$ and $n$ in terms of the distance vectors $\vec{r}_{ij}$. To this end, we first express $\vec{a}_{in}$ in terms of the positions $\vec{h}_{in,m}$ of the vertices that define the polygonal face:
\begin{equation}
  \vec{a}_{in} = \frac{1}{2}\sum_{m=1}^{N_{in}}{\Delta\vec{h}_{in,m}\times\Delta\vec{h}_{in,m+1}}\label{eq:orientedArea}
\end{equation}
with
\begin{equation}
	\Delta\vec{h}_{in,m} = \vec{h}_{in,m} - \vec{r}_i
\end{equation}
being the vertex position relative to the position of cell $i$.
In \eref{eq:orientedArea}, the sum is over all $N_{in}$ vertices of the face, sorted in counter-clockwise order as seen from cell $n$ (right hand rule with the thumb pointing from cell $i$ to cell $n$). The cross $\times$ denotes the vector product and $\vec{h}_{in,N_{in}+1}\equiv\vec{h}_{in,1}$.

It remains to compute the position of a vertex that has the four abutting cells $i,j,l,p$, which we denote by $\vec{h}_{ijlp}$, relative to the position of cell $i$:  $\Delta\vec{h}_{ijlp}=\vec{h}_{ijlp}-\vec{r}_i$.
The relative vertex position in terms of the distance vectors $\vec{r}_{ij}$, $\vec{r}_{il}$, and $\vec{r}_{ip}$ is:
\begin{equation}
  \Delta\vec{h}_{ijlp} = Z_{ijlp}\vec{H}_{ijlp}\text{,}\label{eq:vertexPosition}
\end{equation}
where
\begin{align}
  Z_{ijlp} &= \frac{1}{2[\vec{r}_{ij}\cdot(\vec{r}_{il}\times\vec{r}_{ip})]} \\
  \vec{H}_{ijlp} &= \lvert\vec{r}_{ij}\rvert^2(\vec{r}_{il}\times\vec{r}_{ip}) 
      + \lvert\vec{r}_{il}\rvert^2(\vec{r}_{ip}\times\vec{r}_{ij}) \notag\\
      &\qquad\qquad\qquad\qquad + \lvert\vec{r}_{ip}\rvert^2(\vec{r}_{ij}\times\vec{r}_{il})\text{.}\label{eq:vertexPositionH}
\end{align}

That \esrref{eq:vertexPosition}{eq:vertexPositionH} yield the right vertex position can be seen as follows. The vertex position $\vec{h}_{ijlp}$ can be regarded as the intersection of the three planes that respectively orthogonally bisect the lines between the positions of cells $i$ and $j$, $i$ and $l$, and $i$ and $p$. Any point $\vec{x}$ on each of these planes respectively fulfills
\begin{align}
  0 &= \vec{r}_{ij}\cdot\Big(\bm{x}-[\vec{r}_i+\vec{r}_j]/2\Big)\\
  0 &= \vec{r}_{ik}\cdot\Big(\bm{x}-[\vec{r}_i+\vec{r}_l]/2\Big)\\
  0 &= \vec{r}_{ip}\cdot\Big(\bm{x}-[\vec{r}_i+\vec{r}_p]/2\Big)\text{.}
\end{align}
Insertion verifies that $\vec{x}=\vec{h}_{ijlp}$ as defined by \esrref{eq:vertexPosition}{eq:vertexPositionH} fulfills all three equations.
Note that \esrref{eq:totalEnergy}{eq:vertexPositionH} express the system energy $e$ only in terms of distance vectors $\vec{r}_{ij}$, where $i$ and $j$ are neighboring cells.

\subsection{Forces}
\label{sec:force}
To compute the first derivative of the total energy $e$, we successively use the chain rule of differentiation.
We first compute the derivative of the cell energy $e_i$ with respect to a distance vector $\vec{r}_{ij}$. We have:
\begin{equation}
  \frac{\partial e_i}{\partial \vec{r}_{ij}}
  = 2k_V\big(v_i-1\big)\frac{\partial v_i}{\partial\vec{r}_{ij}}
  +2\big(s_i-s_0\big)\frac{\partial s_i}{\partial\vec{r}_{ij}}\text{.}
\end{equation}
The derivatives of volume and surface are:
\begin{align}
  \frac{\partial v_i}{\partial r_{ij}^\alpha} &= \frac{1}{6}\left(a_{ij}^\alpha + \sum_n{r_{in}^\beta\frac{\partial a_{in}^\beta}{\partial r_{ij}^\alpha}}\right) \\
  \frac{\partial s_i}{\partial r_{ij}^\alpha} &= \sum_n{\frac{a_{in}^\beta}{\lvert\vec{a}_{in}\rvert}\frac{\partial a_{in}^\beta}{\partial r_{ij}^\alpha}} \text{.}
\end{align}
Here, both sums are over all neighbors $k$ of cell $i$.
Moreover, here and in the following, Greek letters represent dimension indices: $\alpha,\beta,\dots\in\lbrace x,y,z\rbrace$ and we use Einstein convention, i.e.\ summation over same indices is implied. 

The derivative of the oriented area is:
\begin{equation}
\begin{aligned}
  \frac{\partial a_{in}^\beta}{\partial r_{ij}^\alpha} &= \frac{\varepsilon^{\beta\gamma\delta}}{2}\sum_{m=1}^{N_{in}}{}\bigg(
    \frac{\partial\Delta h_{in,m}^\gamma}{\partial r_{ij}^\alpha}\Delta h_{in,m+1}^\delta \\
  &\qquad\qquad\qquad\qquad
  + \Delta h_{in,m}^\gamma\frac{\partial\Delta h_{in,m+1}^\delta}{\partial r_{ij}^\alpha}\bigg)\text{,}
\end{aligned}
\end{equation}
where $\varepsilon^{\beta\gamma\delta}$ denotes the Levi-Civita Symbol.

It remains to compute the derivative of the position of a vertex abutting cells $i,j,l,p$ relative to the position of cell $i$, $\Delta\vec{h}_{ijlp}$, with respect to a distance vector $\vec{r}_{ij}$:
\begin{equation}
  \frac{\partial \Delta h_{ijlp}^\eta}{\partial r_{ij}^\alpha} = 
    \frac{\partial Z_{ijlp}}{\partial r_{ij}^\alpha}H_{ijlp}^\eta 
  + Z_{ijlp}\frac{\partial H_{ijlp}^\eta}{\partial r_{ij}^\alpha}
\end{equation}
with
\begin{align}
  \frac{\partial Z_{ijlp}}{\partial r_{ij}^\alpha} &= -2Z_{ijlp}^2\varepsilon^{\alpha\gamma\delta}r_{il}^\gamma r_{ip}^\delta \\
  \frac{\partial H_{ijlp}^\eta}{\partial r_{ij}^\alpha} &= 2r_{ij}^\alpha\varepsilon^{\eta\gamma\delta}r_{il}^\gamma r_{ip}^\delta 
  + \varepsilon^{\alpha\eta\gamma}\Big(\lvert\vec{r}_{il}\rvert^2r_{ip}^\gamma -\lvert\vec{r}_{ip}\rvert^2r_{il}^\gamma\Big)\text{.}
\end{align}

We now know the derivative of a cell energy $e_i$ with respect to the distance to any of its neighbors $j$, $\partial e_i/\partial\vec{r}_{ij}$.  Using the chain rule, the derivative of $e_i$ with respect to neighbor $j$'s position is
\begin{equation}
  \frac{\partial e_i}{\partial\vec{r}_{j}} = \frac{\partial e_i}{\partial\vec{r}_{ij}}\label{eq:enDerivative1}
\end{equation}
and the derivative of the energy of cell $i$ with respect to its own position is
\begin{equation}
  \frac{\partial e_i}{\partial\vec{r}_{i}} = -\sum_j{\frac{\partial e_i}{\partial\vec{r}_{ij}}}\text{.}\label{eq:enDerivative2}
\end{equation}
Here, the sum is over all neighbors $j$ of cell $i$.
The total force on a cell $i$ can be written as $\vec{f}_i=-\sum_k{\partial e_k/\partial\vec{r}_i}$, where the sum is over all cells $k$ of the network. Inserting \esrref{eq:enDerivative1}{eq:enDerivative2}, this becomes
\begin{equation}
  \vec{f}_i = 
  - \sum_j{\frac{\partial e_j}{\partial \vec{r}_{ji}}}  
  + \sum_j{\frac{\partial e_i}{\partial \vec{r}_{ij}}}\text{,}  \label{eq:force}
\end{equation}
where both sums are over all cells $k$ of the network and the second sum is over all neighbors $j$ of cell $i$.

\subsection{Hessian matrix}
\label{sec:hessianMatrix}
We start by deriving the derivative of the energy of a cell $i$ with respect to the cell distance vectors $\vec{r}_{ij}$ where cell $j$ is among the neighbors of cell $i$.
\begin{equation}
\begin{aligned}
  \frac{\partial^2 e_i}{\partial r_{ij}^\alpha\partial r_{ik}^\beta}
  &= 2k_V\frac{\partial v_i}{\partial r_{ij}^\alpha}\frac{\partial v_i}{\partial r_{ik}^\beta} + 2k_V\big(v_i-1\big)\frac{\partial^2 v_i}{\partial r_{ij}^\alpha\partial r_{ik}^\beta} \\
  &\quad + 2\frac{\partial s_i}{\partial r_{ij}^\alpha}\frac{\partial s_i}{\partial r_{ik}^\beta} + 2\big(s_i-s_0\big)\frac{\partial^2 s_i}{\partial r_{ij}^\alpha\partial r_{ik}^\beta}\text{.}
\end{aligned}
\end{equation}
The first derivatives of volume and surface are documented in the previous section.  The second derivatives are:
\begin{align}
  \frac{\partial^2 v_i}{\partial r_{ij}^\alpha\partial r_{ik}^\beta} &= \frac{1}{6}\left(
  \frac{\partial a_{ij}^\alpha}{\partial r_{ik}^\beta}
  +\frac{\partial a_{ik}^\beta}{\partial r_{ij}^\alpha} 
  + \sum_n{r_{in}^\gamma\frac{\partial a_{in}^\gamma}{\partial r_{ij}^\alpha\partial r_{ik}^\beta}}
  \right) \\
  \frac{\partial^2 s_i}{\partial r_{ij}^\alpha\partial r_{ik}^\beta} &= 
  	\sum_n{\frac{1}{\lvert\vec{a}_{in}\rvert}\Bigg[
	  \frac{\partial a_{in}^\gamma}{\partial r_{ij}^\alpha}\frac{\partial a_{in}^\gamma}{\partial r_{ik}^\beta} } \notag\\
      &\qquad\qquad\qquad
	  - \left(\frac{a_{in}^\gamma}{\lvert\vec{a}_{in}\rvert}\frac{\partial a_{in}^\gamma}{\partial r_{ij}^\alpha}\right)
	    \left(\frac{a_{in}^\delta}{\lvert\vec{a}_{in}\rvert}\frac{\partial a_{in}^\delta}{\partial r_{ik}^\beta}\right) \notag\\
      &\qquad\qquad\qquad
      + a_{in}^\gamma\frac{\partial a_{in}^\gamma}{\partial r_{ij}^\alpha\partial r_{ik}^\beta}
	  \Bigg] \text{.}
\end{align}
The second derivative of the oriented area is:
\begin{equation}
\begin{aligned}
  \frac{\partial a_{in}^\gamma}{\partial r_{ij}^\alpha\partial r_{ik}^\beta} &= 
  \frac{\varepsilon^{\beta\gamma\delta}}{2}\sum_{m=1}^{N_{in}}\Bigg(
    \frac{\partial\Delta h_{in,m}^\gamma}{\partial r_{ij}^\alpha} \frac{\partial\Delta h_{in,m+1}^\delta}{\partial r_{ik}^\beta}\\
  &\qquad\qquad\qquad 
  + \frac{\partial\Delta h_{in,m}^\gamma}{\partial r_{ik}^\beta} \frac{\partial\Delta h_{in,m+1}^\delta}{\partial r_{ij}^\alpha} \\
  &\qquad\qquad\qquad 
  + \frac{\partial^2\Delta h_{in,m}^\gamma}{\partial r_{ij}^\alpha\partial r_{ik}^\beta}\Delta h_{in,m+1}^\delta \\
  &\qquad\qquad\qquad 
  + \Delta h_{in,m}^\gamma\frac{\partial^2\Delta h_{in,m+1}^\delta}{\partial r_{ij}^\alpha\partial r_{ik}^\beta}
  \Bigg)\text{.}
\end{aligned}
\end{equation}
The second derivative of the relative vertex position $\Delta\vec{h}_{ijlp}$ is:
\begin{equation}
\begin{aligned}
  \frac{\partial^2 \Delta h_{ijlp}^\eta}{\partial r_{ij}^\alpha\partial r_{ik}^\beta} &= 
  \frac{\partial Z_{ijlp}}{\partial r_{ij}^\alpha}\frac{\partial H_{ijlp}^\eta}{\partial r_{ik}^\beta} 
+ \frac{\partial Z_{ijlp}}{\partial r_{ik}^\beta}\frac{\partial H_{ijlp}^\eta}{\partial r_{ij}^\alpha}  \\
  &\qquad 
  + \frac{\partial^2 Z_{ijlp}}{\partial r_{ij}^\alpha\partial r_{ik}^\beta}H_{ijlp}^\eta 
+ Z_{ijlp}\frac{\partial^2 H_{ijlp}^\eta}{\partial r_{ij}^\alpha\partial r_{ik}^\beta}\text{,}
\end{aligned}
\end{equation}
where cell $k$ is one of $j,l,p$.  For $k=j$, we obtain for the second derivatives of $Z_{ijlp}$ and $\vec{H}_{ijlp}$:
\begin{align}
  \frac{\partial^2 Z_{ijlp}}{\partial r_{ij}^\alpha\partial r_{ij}^\beta} &= \frac{2}{Z_{ijlp}}\frac{\partial Z_{ijlp}}{\partial r_{ij}^\alpha}\frac{\partial Z_{ijlp}}{\partial r_{ij}^\beta} \\
  \frac{\partial^2 H_{ijlp}^\eta}{\partial r_{ij}^\alpha\partial r_{ij}^\beta} &= 2\delta^{\alpha\beta}\varepsilon^{\eta\gamma\delta}r_{il}^\gamma r_{ip}^\delta 
  \text{,}
\end{align}
where $\delta^{\alpha\beta}$ is the Kronecker symbol.
For $k=l$, the derivatives are:
\begin{align}
  \frac{\partial^2 Z_{ijlp}}{\partial r_{ij}^\alpha\partial r_{il}^\beta} &= \frac{2}{Z_{ijlp}}\frac{\partial Z_{ijlp}}{\partial r_{ij}^\alpha}\frac{\partial Z_{ijlp}}{\partial r_{ij}^\beta} 
  -2Z_{ijlp}^2\varepsilon^{\alpha\beta\delta} r_{ip}^\delta\\
  \frac{\partial^2 H_{ijlp}^\eta}{\partial r_{ij}^\alpha\partial r_{il}^\beta} &= 2(r_{il}^\beta\varepsilon^{\alpha\eta\delta}-r_{ij}^\alpha\varepsilon^{\beta\eta\delta})r_{ip}^\delta 
 +\varepsilon^{\alpha\beta\eta}\lvert\vec{r}_{ip}\rvert^2
  \text{.}
\end{align}
Finally, for $k=p$, they are:
\begin{align}
  \frac{\partial^2 Z_{ijlp}}{\partial r_{ij}^\alpha\partial r_{ip}^\beta} &= \frac{2}{Z_{ijlp}}\frac{\partial Z_{ijlp}}{\partial r_{ij}^\alpha}\frac{\partial Z_{ijlp}}{\partial r_{ip}^\beta} 
  +2Z^2\varepsilon^{\alpha\beta\gamma}r_{il}^\gamma\\
  \frac{\partial^2 H_{ijlp}^\eta}{\partial r_{ij}^\alpha\partial r_{ip}^\beta} &= 2(r_{ij}^\alpha\varepsilon^{\beta\eta\gamma}-r_{ip}^\beta\varepsilon^{\alpha\eta\gamma})r_{il}^\gamma - \varepsilon^{\alpha\beta\eta}\lvert\vec{r}_{il}\rvert^2 
  \text{.}
\end{align}

The second derivative of the total energy $e$ with respect to the absolute cell positions $\vec{r}_i$ is the Hessian matrix of the system:
\begin{equation}
  D_{j\alpha,k\beta} = \frac{\partial^2 e}{\partial r_j^\alpha \partial r_k^\beta}\text{.}
\end{equation}
To derive an expression for it, we can proceed analogously to \esrref{eq:enDerivative1}{eq:force}.  Based on the derivatives of the cell energies $e_i$ with respect to relative cell positions $\vec{r}_{ij}$, we obtain:
\begin{equation}
\begin{aligned}
   D_{j\alpha,k\beta} &= 
   \sum_i{\frac{\partial^2 e_i}{\partial r_{ij}^\alpha \partial r_{ik}^\beta}}
  -\sum_l{\frac{\partial^2 e_j}{\partial r_{jl}^\alpha \partial r_{jk}^\beta}} \\
  &\qquad
  -\sum_m{\frac{\partial^2 e_k}{\partial r_{kj}^\alpha \partial r_{km}^\beta}}
  +\delta_{jk}\sum_{l,m}{\frac{\partial^2 e_j}{\partial r_{jl}^\alpha \partial r_{jm}^\beta}}\text{.}
\end{aligned}
\end{equation}
Here, the first sum is over all cells $i$ in the network, the second sum is over all neighbors $l$ of cell $j$, the third sum is over all neighbors $m$ of cell $k$, and the fourth double sum is over all neighbors $l,m$ of cell $j$.

To compute the shear modulus, we also need the second derivatives of the energy with respect to the simple shear variable $\gamma$. We obtain the derivatives involving simple shear using the chain rule and \esrref{eq:skewedPbc}{eq:totalEnergy}:
\begin{equation}
	\frac{\partial^2 e}{\partial\gamma^2} = l_y^2\sum_{i,j,k}{q_{ij}^yq_{ik}^y\frac{\partial^2 e_i}{\partial r_{ij}^x\partial r_{ik}^x}}\text{,}
\end{equation}
where the sum runs over all cells $i$ and all combination of neighbors $j,k$ and $l_y$ is the dimensionless length of the box in $y$ direction.   For the mixed derivative involving the simple shear, we obtain:
\begin{equation}
	\frac{\partial^2e}{\partial\gamma\partial r_j^\alpha} = l_y\left(\sum_{i,k}{q_{ik}^y\frac{\partial^2 e_i}{\partial r_{ik}^x\partial r_{ij}^\alpha}}
    -\sum_{l,k}{q_{jk}^y\frac{\partial^2 e_j}{\partial r_{jk}^x\partial r_{jl}^\alpha}}\right)\text{,}
\end{equation}
where the first sum is over all cells $i$ of the network and all neighbors $k$ of $i$, and the second sum is over all combination of neighbors $l,k$ of cell $j$.

\subsection{Computation of the shear modulus}
\label{sec:shearModulus}
In \sref{sec:energy} we expressed the energy $e$ of the system only in terms of the distance vectors of neighboring cells, $\vec{r}_{ij}$, and in \sref{sec:pbc} we expressed these distance vectors in terms of the absolute position vectors $\vec{r}_i$, and the simple shear variable $\gamma$.  Thus, the energy can be expressed as $e=e(\lbrace\vec{r}_i\rbrace,\gamma)$. 

To introduce the long-time simple shear modulus $g$, we define $e^\mathrm{min}(\gamma)$ as the minimum of $e=e(\lbrace\vec{r}_i\rbrace,\gamma)$ for fixed $\gamma$:
\begin{equation}
	e^\mathrm{min}(\gamma) = \min_{\lbrace\vec{r}_i\rbrace}{e\big(\lbrace \vec{r}_i\rbrace,\gamma\big)}\text{.}
\end{equation}
We denote the set of cell positions with minimal $e$ for given $\gamma$ by $\vec{r}_i^\mathrm{min}(\gamma)$.  Then we define the simple shear modulus as:
\begin{equation}
	g = \frac{1}{N_c}\frac{\d^2 e^\mathrm{min}}{\d\gamma^2}\text{,}\label{eq:shearModulus}
\end{equation}
where $N_c$ is the number of cells and thus the volume of the periodic box in dimensionless units.

The simple shear modulus is directly related to the Hessian matrix of the system.  To derive this relation, we evaluate the derivative in \eref{eq:shearModulus}:
\begin{equation}
	g = \frac{1}{N_c}\left(\frac{\partial^2 e}{\partial\gamma^2} + \sum_{k,\beta}{\frac{\partial^2 e}{\partial r_k^\beta\partial\gamma}\dot{r}_k^{\mathrm{min},\beta}}\right)\label{eq:shearModulus2}
\end{equation}
The second term in the parenthesis can be transformed as follows:
\begin{equation}
	\sum_{k,\beta}{\frac{\partial^2 e}{\partial r_k^\beta\partial\gamma}\dot{r}_k^{\mathrm{min},\beta}} = \sum_m{\left[\sum_{j,\alpha}{\frac{\partial^2 e}{\partial r_j^\alpha\partial\gamma}u_{j\alpha}^m}\right]\left[\sum_{k,\beta}{u_{k\beta}^m\dot{r}_k^{\mathrm{min},\beta}}\right]}\text{,}\label{eq:step1}
\end{equation}
where the sum is over all eigenvalues $\omega_m^2$ of the Hessian, and $u_{j\alpha}^m$ are the corresponding normalized eigenvectors:
\begin{equation}
	D_{j\alpha,k\beta} = \sum_m{\omega_m^2u_{j\alpha}^mu_{k\beta}^m}\text{.}\label{eq:diagonalizationHessian}
\end{equation}
Note that since we minimize with respect to the $\vec{r}^\mathrm{min}$, the Hessian $D_{j\alpha,k\beta}$ is non-negative.
To further transform \eref{eq:step1}, we take the total derivative of the force balance condition $0=\partial e(\lbrace\vec{r}_i^\mathrm{min}(\gamma)\rbrace,\gamma)/\partial r_j^\alpha$ with respect to $\gamma$, which yields:
\begin{equation}
0 = \sum_{k,\beta}{D_{j\alpha,k\beta}\;\dot{r}_k^{\mathrm{min},\beta}} + \frac{\partial^2 e}{\partial r_j^\alpha\partial\gamma}\text{.}\label{eq:differentialForceBalance}
\end{equation}
Here, $\dot{r}_k^{\mathrm{min},\beta}=\d r_k^{\mathrm{min},\beta}/\d\gamma$.  Inserting \eref{eq:diagonalizationHessian} and computing the $3N_c$-dimensional scalar product with $u_{j\alpha}^m$ yields:
\begin{equation}
0 = \omega_m^2\sum_{k,\beta}{u_{k\beta}^m\;\dot{r}_k^{\mathrm{min},\beta}} + \sum_{j,\alpha}{u_{j\alpha}^m\frac{\partial^2 e}{\partial r_j^\alpha\partial\gamma}}\text{.}\label{eq:step2}
\end{equation}
To simplify the sum over $m$ in \eref{eq:step1}, we distinguish between zero modes, for which $\omega_m=0$ and non-zero modes with $\omega_m>0$.  For zero modes $m$, the second term in \eref{eq:step2} has to be zero.  Because this is the same term as the first factor in the sum over $m$ in \eref{eq:step1}, zero modes do not contribute to this sum.
For the non-zero modes, we can substitute \eref{eq:step2} into \eref{eq:step1} and obtain from \eref{eq:shearModulus2}:
\begin{equation}
	g = \frac{1}{N_c}\left(\frac{\partial^2 e}{\partial\gamma^2}-\sum_m{\frac{1}{\omega_m^2}\left[\sum_{j,\alpha}{\frac{\partial^2e}{\partial\gamma\partial r_j^\alpha}u_{j\alpha}^m}\right]^2}\right)\text{.}\label{eq:shearModulusComputation}
\end{equation}
In this equation, the sum over $m$ excludes zero modes of the Hessian $D_{j\alpha,k\beta}$.

\subsection{Alternative computation of the shear modulus using the extended Hessian}
\label{sec:extendedHessian}
Alternatively to \eref{eq:shearModulusComputation}, the shear modulus can also be computed from directly the eigen spectrum of the extended Hessian
\begin{equation}
  \bar{D}_{pq} = \frac{\partial^2 e}{\partial z_p\partial z_q}\text{,}
\end{equation}
where $(z_p)$ is a $(3N_c+1)$-dimensional vector comprising all cell positions and the shear degree of freedom $\gamma$: $(z_p)=(\vec{r}_1, \dots, \vec{r}_{N_c}, \gamma)$.   The eigenvalues of $\bar{D}_{pq}$ are $\bar{\omega}_m^2$ and the normalized eigenvectors are $\bar{u}_p^m$:
\begin{equation}
  \bar{D}_{pq} = \sum_m{\bar{\omega}_m^2\bar{u}_p^m\bar{u}_q^m}\text{.}
\end{equation}
Note that for shear-stabilized states, the extended Hessian $\bar{D}_{pq}$ is non-negative.

To derive an alternative formula for the shear modulus, we combine \esref{eq:shearModulus2} and \seref{eq:differentialForceBalance} into:
\begin{equation}
	N_c g \delta_{\gamma p} = \bar{D}_{pq}\dot{z}_q^\mathrm{min}
\end{equation}
with $\delta$ being the Kronecker delta and $(\dot{z}_q^\mathrm{min})=(\dot{\vec{r}}_1^\mathrm{min}, \dots, \dot{\vec{r}}_{N_c}^\mathrm{min}, 1)$.
The $(3N_c+1)$-dimensional scalar product with an eigenvector $\bar{u}_p^m$ yields:
\begin{equation}
	N_c g \bar{u}_\gamma^m = \bar{\omega}_m^2\sum_q{\bar{u}_q^m\dot{z}_q^\mathrm{min}}\label{eq:alt_step1}
\end{equation}
for each eigenvalue $m$.

To evaluate the shear modulus, we distinguish two cases.  First, whenever there exists a zero mode $m$ (i.e.\ $\bar{\omega}_m=0$) with nonzero shear component $\bar{u}_\gamma^m$, then \eref{eq:alt_step1} implies that the shear modulus $g$ has to vanish.

Second, all zero modes $m$ have vanishing shear components: $\bar{u}_\gamma^m=0$.  In this case, the following relation
\begin{equation}
	\delta_{\gamma q} = \sum_m{\bar{u}_\gamma^m\bar{u}_q^m}
\end{equation}
holds also when only summing over non-zero modes $m$.
As a consequence, the following relation holds:
\begin{equation}
	1 = \sum_{m,q}{\bar{u}_\gamma^m\bar{u}_q^m\dot{z}_q^\mathrm{min}}\text{,}
\end{equation}
where we sum again only over non-zero modes $m$, because $\dot{z}_\gamma^\mathrm{min}=1$.
Insertion of \eref{eq:alt_step1} finally yields:
\begin{equation}
	g = \frac{1}{N_c}\left[\sum_{m}{\frac{(\bar{u}_\gamma^m)^2}{\bar{\omega}_m^2}}\right]^{-1}\text{.}\label{eq:extHessianShearModulus}
\end{equation}
Here, the sum is over all non-zero modes $m$.

\section{Numerical implementation of the 3D Voronoi model}
\label{sec:implementation}
\subsection{Initial conditions and parameter values}
\label{sec:parameters}
For all simulations, the cells are initially assigned independent random positions $\vec{r}_i$ drawn from a uniform distribution.
To generate the Voronoi tessellations for a given set of cell positions $\lbrace\vec{r}_i\rbrace$, we used the \verb voro++  library by Chris Rycroft (version 0.4.6, \cite{Rycroft2009}).  Cell surface areas and volumes, forces, and the shear modulus were then computed as described in \ssref{sec:pbc}--\ref{sec:shearModulus}.  In particular, shear moduli $g$ were computed using a cutoff value of $10^{-14}$ below which eigenvalues of the Hessian were regarded as zero modes and thus disregarded for the sum to compute $g$.  We checked that $g$ was largely independent of this cutoff over a range of cutoff values.  To diagonalize the Hessian, we used the \verb Eigen3  library (version 3.2.7, \footnote{Available at: \protect\url{http://eigen.tuxfamily.org}}).

We studied the cases $k_V=10^{-3},10^{-2},\dots,10^3$ and $s_0$ in the range $[0,5.9]$ in steps of $0.1$ and in the range $[5.35,5.45]$ in steps of $10^{-3}$.  For each parameter pair $(s_0,k_V)$, we ran 100 minimizations each of which was initialized with a different set of random cell positions.  The system size was $N_c=512$ unless stated otherwise.

\subsection{Numerical energy minimization}
\label{sec:energyMinimization}
\begin{table}
  \begin{tabular}{p{6cm}|r}
    \bf Quantity & \bf Value \\\hline
    Initial step size, \verb step_size  & $0.01$ \\
    Line minimization tolerance, \verb tol  & $0.01$ \\
    Total force norm cutoff, \verb epsabs  & $10^{-12}\sqrt{N_\mathrm{dof}}$ \\
    Maximal number of iteration steps per minimization & $100N_\mathrm{dof}$
  \end{tabular}
  \caption{Parameter values used for the multidimensional energy minimization.  $N_\mathrm{dof}$ is the number of degrees of freedom varied during the minimization.  For any minimization, all cell positions are included.  Thus, depending on whether the shear degree of freedom is also included, $N_\mathrm{dof}=3N_c$ or $N_\mathrm{dof}=3N_c+1$.\label{tab:minParameters}}
\end{table}
To minimize the energy of the system, we used the BFGS2 multidimensional minimization routine of the GNU Scientific Library (GSL, version 2.1, \footnote{Available at: \protect\url{https://www.gnu.org/software/gsl/}}) \cite{Fletcher2000}.  The parameters used for one GSL minimization are listed in \tref{tab:minParameters}, where $N_\mathrm{dof}$ is the number of degrees of freedom varied during the minimization.  We tested that different values for the individual parameters did not improve the minimization, i.e.\ the norm of the total force vector after the minimization was not smaller for different parameter values.

Often, the GSL library could not further minimize the energy and did not reach the total force cutoff listed in \tref{tab:minParameters}.  In these cases, we tested whether the total force norm was at least below a cutoff of $C=10^{-7}\sqrt{N_\mathrm{dof}}$.  If that was not the case, we started another GSL minimization starting with the last set of cell positions.  We repeated GSL minimizations until the cutoff $C$ was reached or 10 GSL minimizations had been performed.  

To obtain shear-stabilized force-balanced states, we first ran up to 10 GSL minimizations varying all $N_c$ cell positions $\vec{r}_i$, i.e.\ $N_\mathrm{dof}=3N_c$.  Afterwards, to shear-stabilize the system, we included the shear degree of freedom $\gamma$ into the minimization, simultaneously varying $N_\mathrm{dof}=3N_c+1$ degrees of freedom, running again up to 10 GSL minimizations.  We discarded all simulation runs that had a total force norm larger than $10^{-4}$ after the minimization procedure or a negative shear modulus smaller than $g<-10^{-5}$.  An exception are \fsref{fig:modelAndTransition}a and \sfref{fig:residualstressesAndMinSurface}b, where we needed to increase the force cutoff to $10^{-3}$. This is because for large $k_V$ deep in the solid regime, we couldn't minimize the total force below $10^{-4}$ in many cases.

\subsection{Probing the solid vicinity of the transition point}
\label{sec:solidVicinity}
We performed dedicated simulations to explore the solid vicinity of the transition point.  To this end, we first created an energy-minimized at its transition point and then decreased $s_0$ using exponentially increasing steps.

To tune a configuration right at its transition point, we used bisection on the $s_0$ parameter with the initial left and right bracket values of $s_0=5.38$ and $s_0=5.44$.  In each bisection step, the $s_0$ value is set to the average of the current bracket values and the energy is minimized as described in the previous section.  If the new state is solid, it is kept for the next minimization, and the next left bracket value is set to the current $s_0$ value.  However, if the state is fluid, the system is reverted to the last solid state (corresponding to the left bracket value) and the next right bracket value is set to the current $s_0$ value.  
We choose to keep the solid but not the fluid states to reduce the probability to switch the ``inherent state'' during the bisection.
As another measure to avoid switching the ``inherent state'', we included a check verifying that whenever the energy-minimized state at the current $s_0$ value is solid, the system at the right bracket is still fluid if we start the minimization from the energy-minimized state at the current $s_0$ value.
We performed 13 such bisection steps, and a configuration was deemed solid if its shear modulus was larger than $10^{-7}$.  

Once we obtained an energy-minimized state at its rigidity transition, we explored the solid regime by iteratively reducing $s_0$ in exponentially growing steps, each time minimizing the energy as described in the previous section.
As a final measure to exclude simulation runs where the ``inherent state'' changed, we computed the mean-squared deviations $R^2$ of the cell positions $\vec{r}_i$ at some point $s_0$ in the solid regime from the cell positions at the transition point $s_0^\ast$ (corrected by any overall translation).  To exclude simulations where the ``inhered state'' changed, we made sure $R^2$ was not too large in our simulations.  More precisely, we excluded simulations with $R^2>100\delta s_0^2$ for any of the $s_0$ probed when exploring the solid vicinity of $s_0^\ast$.

\section{Additional numerical analysis of the 3D Voronoi model}
\subsection{Periodic Voronoi packings}
\label{sec:periodicPackings}
The average cell surface areas $\langle s\rangle$ for Kelvin ($\langle s\rangle\approx5.306$) and Weaire-Phelan packings ($\langle s\rangle\approx5.288$) are well-known.  However, these figures correspond to configurations with curved cell outlines.

To numerically obtain the $\langle s\rangle$ values for the Voronoi packings corresponding to the Kelvin and Weaire-Phelan structures, we proceeded as follows.  For the Kelvin Voronoi structure, we prepared a cubic periodic box with side length $2^{1/3}$ with two cells at $\vec{r}_1=(0,0,0)$ and $\vec{r}_2=2^{-2/3}(1,1,1)$ and minimized the total surface area.  This initial configuration already corresponded to a minimal surface area with $\langle s\rangle\approx5.315$ and equal volume of both cells.  

\begin{table}
  \begin{tabular}{p{4.7cm}|c}
    \bf Type of cellular polyhedron & \bf Position \\\hline
    \multirow{6}{3cm}{6 cells with 14 faces}
      & $(0.0, 1.5, 1.5)$ \\
      & $(1.0, 1.5, 1.5)$ \\
      & $(0.5, 0.0, 0.5)$ \\
      & $(0.5, 1.0, 0.5)$ \\
      & $(1.5, 0.5, 0.0)$ \\
      & $(1.5, 0.5, 1.0)$ \\\hline
      \multirow{2}{3cm}{2 cells with 12 faces}
      & $(0.5, 0.5, 1.5)$ \\
      & $(1.5, 1.5, 0.5)$
  \end{tabular}
  \caption{Initial cell positions to obtain the surface area corresponding to the Weaire-Phelan Voronoi structure.  These cells are put into a in a cubic periodic box with side length 2.\label{tab:WeairePhelanVoronoi}}
\end{table}
For the Weaire-Phelan Voronoi structure, we prepared a state with a cubic periodic box with side length of $2$ and 8 cells at the positions listed in \tref{tab:WeairePhelanVoronoi}.  Without constraining the volume, the surface was already minimal for these initial positions with $\langle s\rangle\approx5.295$.  Note however that the volumes of the cells were slightly different.  All 14-faced cells had $v_i\approx1.0078$ while the two 12-faced cells had $v_i\approx0.9765$.  We tried to constrain all cell volumes to be equal but did not succeed.  We thus compare our result for the disordered transition point only to the Kelvin structure in the main text.

\begin{figure}
  \includegraphics{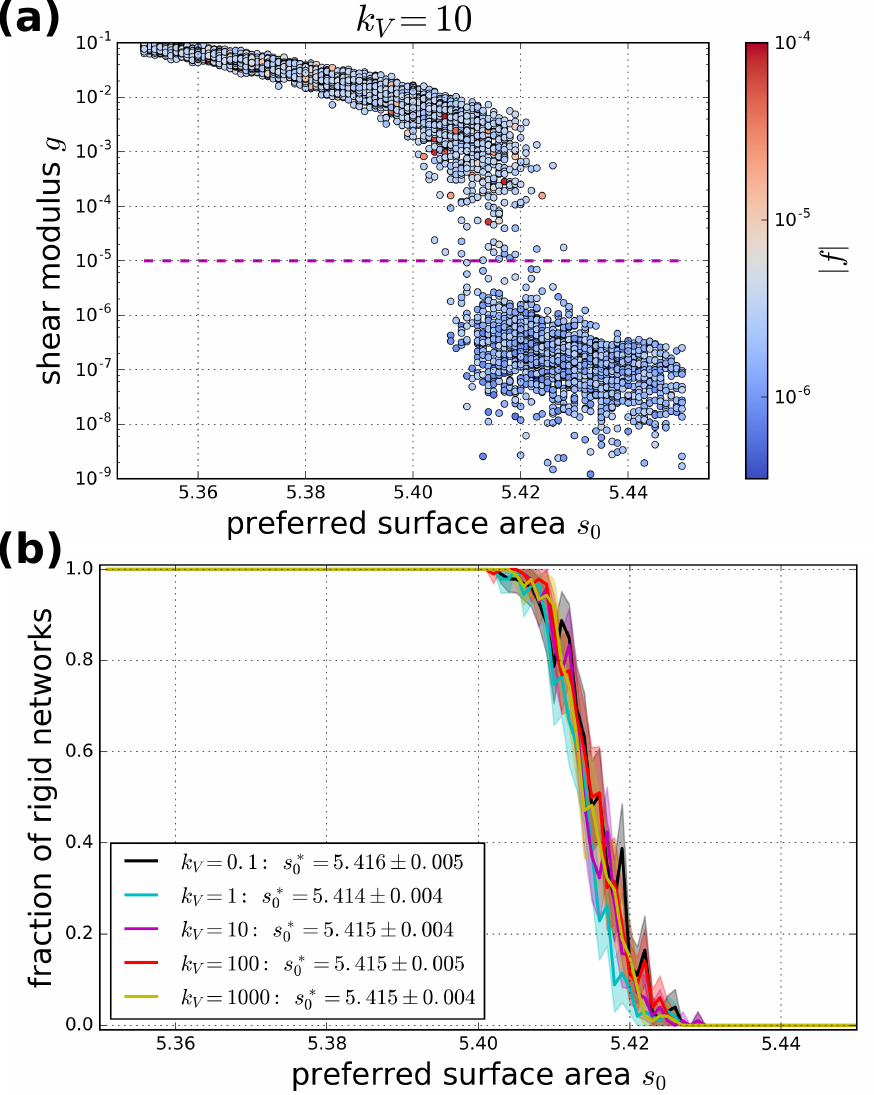}
  \caption{Determination of the precise transition point $s_0^\ast$.  \textbf{(a)} Scatter plot of the shear modulus $g$ over the preferred surface area $s_0$ for $k_V=10$ in order to determine a cutoff value on $g$ (magenta dashed line) to define a network as rigid.  Note that we chose different cutoff values for different values of $k_V$ (plots are similar for other values of $k_V$).  The color of the dots indicate the norm of the total force vector at the end of the minimization, $\lvert f\rvert$. \textbf{(b)} Fraction of rigid networks depending on the preferred surface area $s_0$.  The shaded regions indicate the respective uncertainties, computed as twice the standard deviation of the corresponding binomial distribution. The legend lists the resulting values for average and standard deviation of the transition point.  System size $N_c=512$.\label{fig:preciseTransitionPoint}}
\end{figure}
\subsection{Determination of the precise transition point}
\label{sec:transitionPoint}
In order to determine the precise transition point, we first quantified the fraction $F$ of rigid networks for different values of the preferred surface area $s_0$.  In the limit of an infinite number of simulation runs, the function $F(s_0)$ can be regarded as the integrated probability distribution of transition points $P(s_0^\ast)$:
\begin{equation}
  F(s_0) = 1 - \int_{-\infty}^{s_0}{P(s_0^\ast)\,\d s_0^\ast}\text{,}
\end{equation}
or conversely, $P(s_0^\ast)=-F'(s_0^\ast)$.  The first two moments of the transition point $\langle s_0^\ast\rangle=\int_{-\infty}^\infty{s_0^\ast P(s_0^\ast)\,\d s_0^\ast}$ and $\langle (s_0^\ast)^2\rangle=\int_{-\infty}^\infty{(s_0^\ast)^2 P(s_0^\ast)\,\d s_0^\ast}$ can be evaluated using partial integration.  Assuming that the transition point is always non-negative, we obtain:
\begin{align}
  \langle s_0^\ast\rangle &= \int_0^\infty{F(s_0)\,\d s_0} \label{eq:tpAvg}\\
  \sigma^2(s_0^\ast) &= 2\int_0^\infty{s_0F(s_0)\,\d s_0} - \langle s_0^\ast\rangle^2 \label{eq:tpVar}\text{,}
\end{align}
where $\sigma(s_0^\ast)$ denotes the standard deviation of the distribution of transition points. 

We defined an energy-minimized network as rigid whenever its shear modulus $g$ was larger than a cutoff value.  To determine this cutoff value for a given $k_V$, we plotted the shear modulus $g$ for each simulation run over $s_0$ (\fref{fig:preciseTransitionPoint}a).  In such plots, we find a clear separation of at least a decade between two clusters of networks.  We interpret all networks belonging to the respective lower cluster as non-rigid and those belonging to the upper cluster as rigid.  For $k_V=10$, our chosen cutoff is indicated by a magenta dashed line in \fref{fig:preciseTransitionPoint}a.  For the cases $k_V=0.001$ and $k_V=0.01$, both clusters were not clearly enough separated to define a sensible cutoff value.  We have thus ignored these parameter values throughout this article.

The resulting values obtained for average $\langle s_0^\ast\rangle$ and standard deviation $\sigma(s_0^\ast)$ of the transition points are listed in \fref{fig:preciseTransitionPoint}b for the different values of $k_V$.  The values for the average transition point range from $\langle s_0^\ast\rangle\approx5.414$ to $\langle s_0^\ast\rangle\approx5.416$.  When respectively varying the cutoff on $g$ between the two clusters, the value of $\langle s_0^\ast\rangle$ varied by up to $\sim10^{-3}$.  Thus, the average transition points for the different values of $k_V$ are not significantly different from each other.

\begin{figure}
    \includegraphics{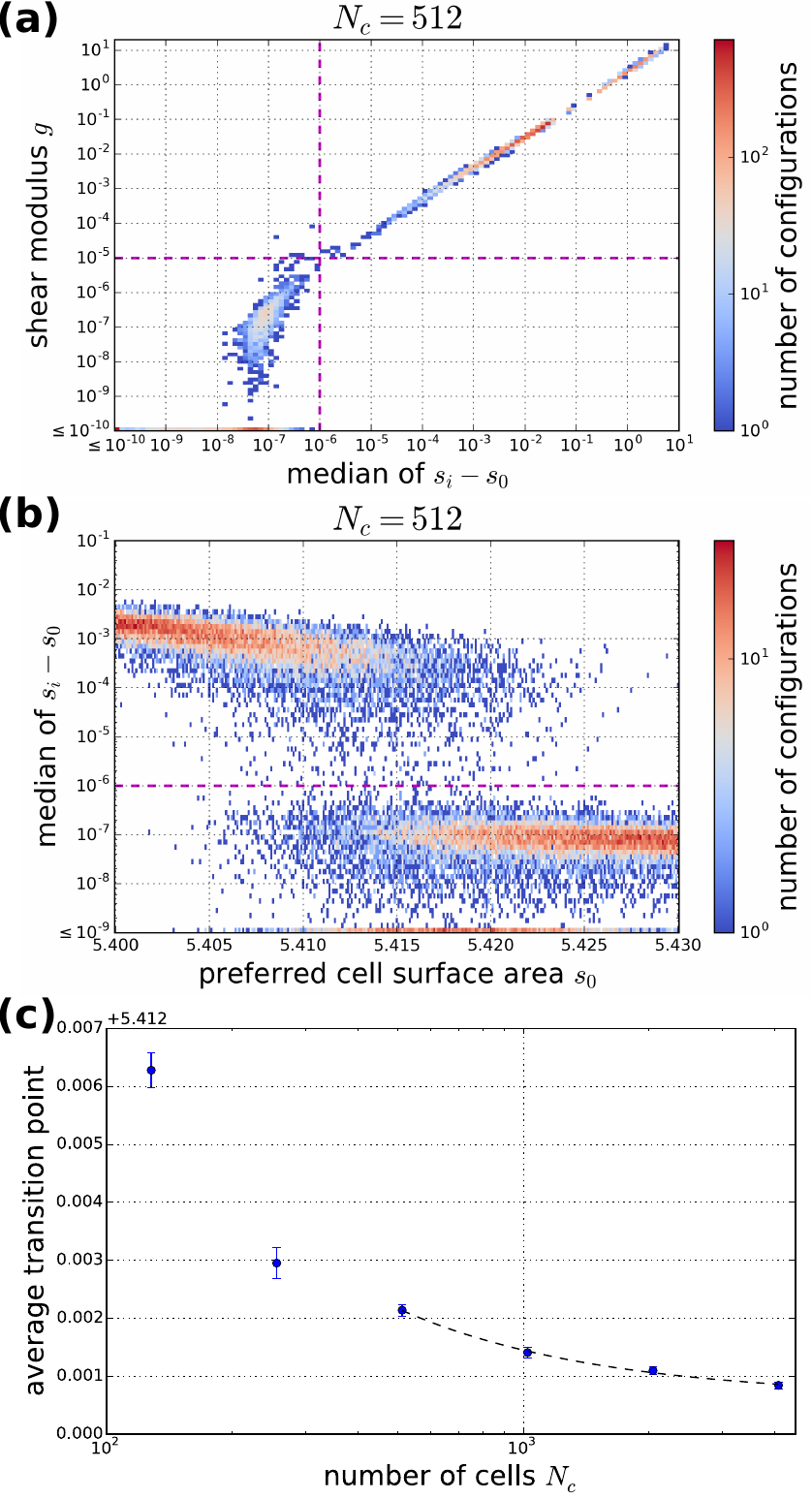}
    \caption{Finite-size scaling of the transition point distribution using the median of $s_i-s_0$.  
    \textbf{(a)} The median of $s_i-s_0$ is a very good indicator for rigidity as measured by the shear modulus $g$.  Magenta lines indicate cutoff values.
    \textbf{(b)} Determination of the cutoff value for the median based on histograms of the median of $s_i-s_0$ depending on $s_0$ with $k_V=10$.
    \textbf{(c)} Average of the transition point distribution $\langle s_0^\ast\rangle$ depending on system size $N_c$.  The black dashed line indicates a power law fit with offset $s_0^\ast\approx5.413$ and exponent $-0.9\pm0.4$, where we only used $N_c\geq512$, because a common fit through all data points was not consistent with the $N_c=256$ data point. 
    \label{fig:median}}
\end{figure}
\subsection{Finite-size scaling}
\label{sec:fss}
We studied the behavior of the transition point distribution for varying system size.  However, to compute the shear modulus for very large systems, we would have needed to diagonalize very big Hessians of size $N_\mathrm{dof}\times N_\mathrm{dof}$, which quickly exhausted the memory of our machines.  We thus chose an alternative approach, where we first realized that the shear modulus correlated very well with the median surface tension in the system (\fref{fig:median}a).  Note in particular that this is the case even for the exceptions visible in \fref{fig:residualstressesAndMinSurface}a and discussed in \sref{sec:exceptions} below.  The cutoff for the median, $10^{-6}$, was chosen such that for each system size $N_c$ it clearly separated the two clusters that appear in the histograms showing the distribution of medians depending on $s_0$ (\fref{fig:median}b).

Using this median cutoff, we extracted the functions $F(s_0)$ for each $N_c$ for $k_V=10$.  Using \esref{eq:tpAvg} and \seref{eq:tpVar}, we then computed the average (\fref{fig:median}c) and variance (\fref{fig:modelAndTransition}b inset) of the transition point distribution.  For the average, we find a scaling exponent of $-0.9\pm0.4$ and a limit value of $s_0^\ast(N_c\rightarrow\infty)=5.413\pm 0.001$.  For the variance, we find a scaling exponent of $-0.90\pm0.04$.

In \fref{fig:modelAndTransition}b, the transition point distributions have been computed based on the $F(s_0)$ function. To numerically compute the derivative while suppressing noise, for each system size $N_c$, we convoluted $F$ with the derivative of a Gaussian with a standard deviation of $\sigma(s_0^\ast)/2$.

\begin{figure}
  \centering
  \includegraphics{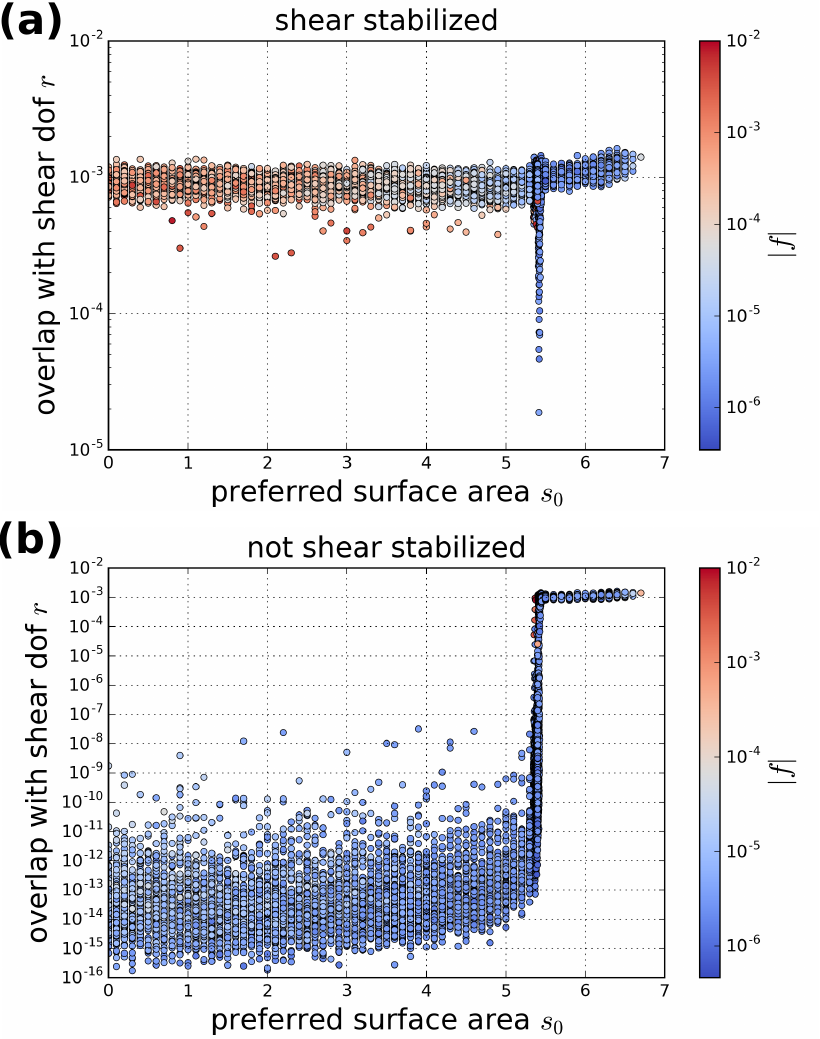}
  \caption{Overlap of the infinitesimal zero modes with the shear degree of freedom for different values of $s_0$ and $k_V=10$ (plots for other values of $k_V$ are similar).  Each dot represents a single energy-minimized network configuration that is \textbf{(a)} shear stabilized or \textbf{(b)} not shear stabilized.  
  In the shear-stabilized case, the overlap in the solid regime is finite, which indicates that residual stresses are necessary to create rigidity.
  To distinguish whether the overlaps that we measure in (a) are finite because of the physics or because of numerical noise, we also plotted the not shear-stabilized case in (b).  We expect the overlap in this case to be zero in the solid regime and indeed, we measure values that are several decades smaller than in the shear-stabilized case (a).
  The color of the dots represents the norm of the $(3N_c+1)$-dimensional (a) or $(3N_c)$-dimensional (b) total force vector after the minimization.
  \label{fig:overlap}}
\end{figure}
\subsection{Residual stresses were necessary to create rigidity.}
\label{sec:overlap}
In the main text, we discuss the full extended Hessian $\bar{D}_{pq}$ and find that the system is rigid below the transition point $s_0^\ast$. 
To numerically check whether residual stresses are necessary to create such rigidity, we study an \emph{unstressed} extended Hessian $\bar{D}_{pq}'$ where the terms that depend on the residual stresses have been removed:
\begin{equation}
  \bar{D}_{pq}' = 2\sum_i{}\Bigg[\frac{\partial s_i}{\partial z_p}\frac{\partial s_i}{\partial z_q} + k_V\frac{\partial v_i}{\partial z_p}\frac{\partial v_i}{\partial z_q}\Bigg]\text{.}
\end{equation}
Here $(z_p)$ is a $(3N_c+1)$-dimensional vector of the cell positions and the shear degree of freedom. 

$\bar{D}_{pq}'$ corresponds to the Hessian of a system with a modified energy functional where we set the preferred surface area of each cell to its actual surface area and the preferred volume of each cell to its actual volume.  This modified system thus corresponds to the original system without residual stresses (cf.\ \eref{eq:dynamicalMatrix} in the main text).

If (and only if) $\bar{D}_{pq}'$ has a zero mode with a nonzero shear component, then the shear modulus of the modified system is zero (\aref{sec:extendedHessian}).  Therefore, if we can show that $\bar{D}_{pq}'$ has such zero modes in the solid phase, we can conclude that residual stresses are necessary to rigidify the original system $\bar{D}_{pq}$.

The zero modes of $\bar{D}_{pq}'$ correspond to the infinitesimal zero modes of the system.  Infinitesimal zero modes are $(3N_c+1)$-dimensional vectors $\vec{w}$ in the kernel of the so-called compatibility matrix $\mathrm{C}$ \cite{Lubensky2015}, i.e.\ $\mathrm{C}\cdot\vec{w}=0$.  For our system, $\mathrm{C}$ is a $(2N_c)\times(3N_c+1)$ matrix defined as:
\begin{equation}
	\mathrm{C} = \begin{pmatrix} 
    \partial s_1/\partial z_1 & \dots & \partial s_1/\partial z_{3N_c+1} \\ 
    & \dots & \\
    \partial s_{N_c}/\partial z_1 & \dots & \partial s_{N_c}/\partial z_{3N_c+1} \\
    \partial v_1/\partial z_1 & \dots & \partial v_1/\partial z_{3N_c+1} \\ 
    & \dots &\\
    \partial v_{N_c}/\partial z_1 & \dots & \partial v_{N_c}/\partial z_{3N_c+1}
    \end{pmatrix}\text{.}
\end{equation}
Note that $\mathrm{C}$ is the same for both original and modified system.
Hence, an infinitesimal zero mode corresponds to a collective change of cell center positions $\lbrace \vec{r}_i\rbrace$ and shear variable $\gamma$ that leaves all cell surfaces and volumes constant to linear order.  

The compatibility matrix has a given number $N_0$ of independent infinitesimal zero modes.  More precisely, this means that there is a set of $N_0$ infinitesimal zero modes $\vec{w}^q$ with $q=1,\dots,N_0$ and $\mathrm{C}\cdot\vec{w}^q=0$, which are orthonormal: $\vec{w}^p\cdot\vec{w}^q=\delta_{pq}$.  We extracted such an orthonormal set of zero modes $\vec{w}^q$ using singular value decomposition of $\mathrm{C}$.

As a tool to study the shear components $w^q_\gamma$ of all infinitesimal zero modes $\vec{w}^q$ at once, we define the overlap $r$ of the kernel of $\mathrm{C}$ with the shear degree of freedom by:
\begin{equation}
  r = \sum_{q=1}^{N_0}{\big(w^q_\gamma\big)^2}\text{.}\label{eq:defOverlap}
\end{equation}
By definition, $0\leq r\leq 1$.   In particular, $r$ is nonzero if and only if there is an infinitesimal zero mode that has a finite shear component.  In other words, $r$ is nonzero if and only if the shear modulus of the modified system is zero.

In \fref{fig:overlap}a, we plot the overlap $r$ depending on the preferred cell surface area $s_0$ for $k_V=10$, where each dot represents an energy-minimized and shear-stabilized state.  We find that the overlap is finite and on the order of $r\sim 10^{-3}$ both in the fluid regime ($s_0>s_0^\ast$) and in the solid regime ($s_0<s_0^\ast$), which holds independent of the value of $k_V$ (data not shown). This shows that the modified system always has zero shear modulus and thus residual stresses are indeed necessary to rigidify the original system.  

Note that close to the transition point $s_0^\ast\approx5.4$, the overlap is occasionally smaller than $10^{-3}$.  However, this occurs only in a fraction of the cases at $s_0^\ast$.  Moreover, the overlap was always larger than $10^{-5}$.
To ensure that the observed values of $r\in [ 10^{-5}, 10^{-3} ]$ in shear-stabilized systems are demonstrably different from zero, we measure the numerical noise in a system where we know the overlap $r$ should be zero. We expect that configurations that have not been shear-stabilized will generically have a finite shear stress in the solid phase, and as we discuss below, a finite shear stress induces a zero overlap $r$. Therefore, we plot the overlap for non-shear-stabilized configurations with $k_V=10$, as shown in \fref{fig:overlap}b.   We find that the measured overlap in the solid phase is always below $r\sim10^{-7}$ and is typically on the order of $r\sim10^{-14}$, which is significantly smaller than $r\in [ 10^{-5}, 10^{-3} ]$.  This supports our interpretation that the overlap is indeed finite in the shear-stabilized systems.

To see why the overlap is zero for non-shear-stabilized states with finite shear stress $\sigma_{xy}$, we first consider the force balance condition combined with the definition for the shear stress:
\begin{equation}
	\frac{\partial e}{\partial z_p} = N_c\sigma_{xy}\delta_{p\gamma}\text{.}
\end{equation}
Here, $\delta$ denotes the Kronecker delta and we use dimensionless units so that $N_c$ corresponds to the system volume.  Using chain rule on the left-hand side and using vector notation, this transforms into:
\begin{equation}
	2\begin{pmatrix} 
    s_1-s_0 \\ 
    \dots \\
    s_{N_c}-s_0 \\
    k_V(v_1-1) \\ 
    \dots \\
    k_V(v_{N_c}-1)
    \end{pmatrix}
    \cdot\mathrm{C} = N_c\sigma_{xy}\vec{\hat{\gamma}}\text{.}
    \label{eq:gammaHat}
\end{equation}
Here, $\vec{\hat{\gamma}}$ is the $(3N_c+1)$-dimensional vector $\vec{\hat{\gamma}}=(0,\dots,0,1)$.
For $\sigma_{xy}\neq0$, we obtain an expression for $\vec{\hat{\gamma}}$ from \eref{eq:gammaHat}, and insertion into $w^q_\gamma=\vec{\hat{\gamma}}\cdot\vec{w}^q$ yields indeed $w^q_\gamma=0$, because $\mathrm{C}\cdot\vec{w}^q=0$.  Thus, the overlap is generally zero in the not shear-stabilized case: $r=0$.
Intuitively, in order for the system to support a finite shear stress, there can be no infinitesimal shear mode that involves the shear degree of freedom, i.e.\ there can be no collective displacement that leaves all surfaces and volumes constant to linear order while also shearing the system.

\begin{figure}
	\centering
 	\includegraphics{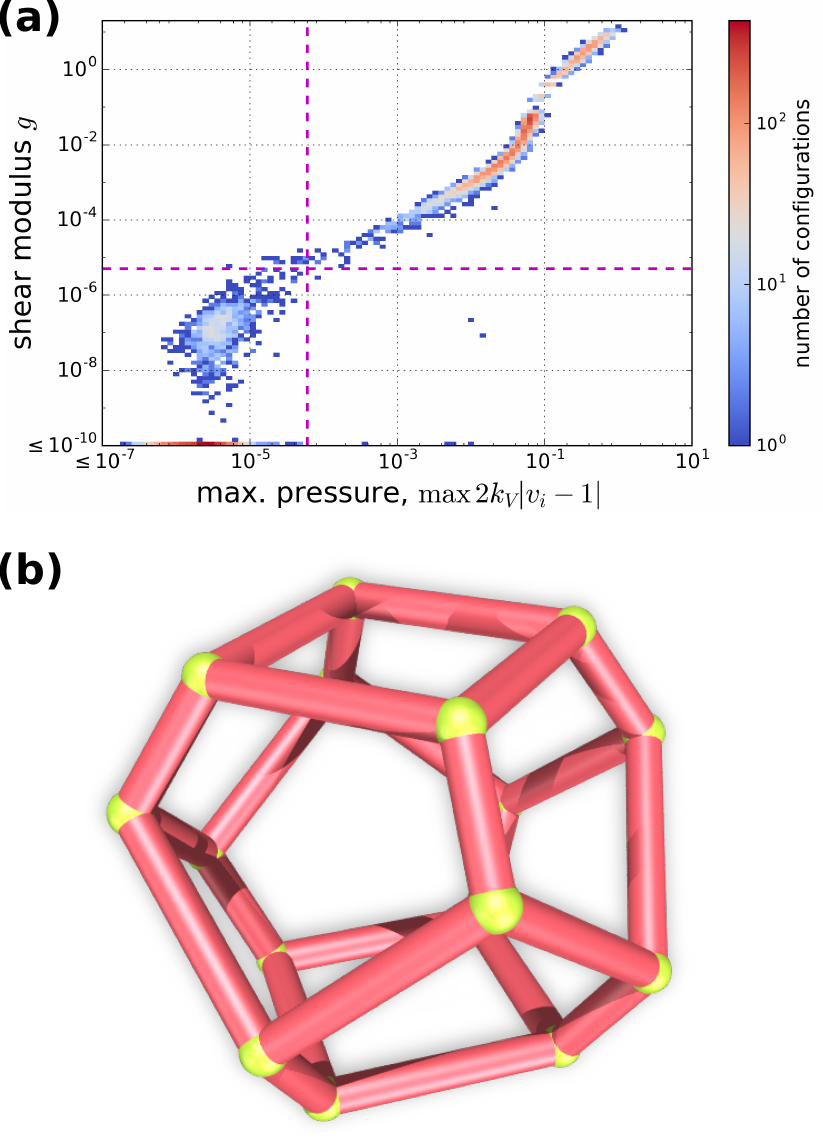}
    \caption{ 
    \textbf{(a)} Relation between rigidity and the existence of residual stresses: Two-dimensional histogram of shear modulus $g$ and maximal magnitude of cell pressure $\mathrm{max}\,2k_V\lvert v_i - 1\rvert$ for $k_V=1$ (cf.\ \fref{fig:residualstressesAndMinSurface}a; plots for different values of $k_V$ are similar). 
    Note that all values of $g$ that are smaller than $10^{-10}$ are mapped onto the $g=10^{-10}$ line, and similarly for $\mathrm{max}\,2k_V\lvert v_i - 1\rvert$.
    \textbf{(b)} Fluid configurations in which residual stresses occur typically contain a truncated square trapezohedron (TST). A TST consists of two opposing squares, each of which is connected to four irregular pentagons, totaling to eight pentagons with minimal surface-to-volume ratio $s_\mathrm{TST}\approx5.4436$.
    \label{fig:g-residualStresses}}
\end{figure}
\subsection{Residual stresses were in most cases sufficient to create rigidity.}
\label{sec:residualStressesCreateRigidity}
In the previous section, we have shown that residual stresses were necessary to create rigidity in all our simulations.  Here, we provide evidence that, at least in the vast majority of simulations, they were also sufficient to create rigidity.


To relate residual stresses to rigidity, we sorted all energy-minimized configurations into two-dimensional histograms with respect to the shear modulus $g$ and with respect to the maximal surface tension magnitude $\mathrm{max}\,2\lvert s_i - s_0\rvert$ (\fref{fig:residualstressesAndMinSurface}a in main text) and maximal pressure magnitude $\mathrm{max}\,2k_V\lvert v_i - 1\rvert$ (\fref{fig:g-residualStresses}a). 
In addition, we marked in each case the cutoff defined in \sref{sec:transitionPoint} below which we interpret a configuration as fluid by a horizontal magenta dashed line (compare \fref{fig:preciseTransitionPoint}).

In both plots, we find one cluster that clearly lies in the fluid regime and several clusters appearing in the solid regime.  We think that the appearance of more than one solid cluster is merely due to our sampling of the $s_0$ values:  Close to the transition point, we varied $s_0$ in small steps of $10^{-3}$ while some distance $\Delta s_0\sim 10^{-1}$ away from the transition point, we varied $s_0$ in steps of $0.1$ (compare \sref{sec:energyMinimization}).

We realized that the fluid cluster in both plots can be separated from the solid clusters not only by the horizontal line, but also by a vertical line, i.e.\ by a cutoff on $\mathrm{max}\,2\lvert s_i - s_0\rvert$ or $\mathrm{max}\,2k_V\lvert v_i - 1\rvert$.  We interpret configurations right of this line as having residual stresses of the respective kind, while configurations left of this line have no residual stresses.
The fact that the lower-right of the quadrants formed by the two lines in both plots is mostly devoid of configurations indicates that residual stresses were in the vast majority of cases sufficient to create rigidity in our simulations.
Finally, although we present here only the plots for the case $k_V=1$  (\fsref{fig:residualstressesAndMinSurface}a and \sfref{fig:g-residualStresses}a), we verified using analogous plots that our conclusions are unchanged for other values of $k_V$ between $0.1$ and $1000$.

\subsubsection{Rare cases of fluid configurations with residual stresses}
\label{sec:exceptions}
To study the rare exception cases of fluid configurations with residual stresses, we ran dedicated sets of simulations for $k_V=1$ and $s_0=5.35\dots5.45$ in steps of $10^{-3}$ where we additionally stored all cell positions of the resulting shear-stabilized energy-minimized states.  Like in \fsref{fig:residualstressesAndMinSurface}a and \sfref{fig:g-residualStresses}a, we observed again few simulation runs that were in the fluid regime while containing significant residual stresses.  Analyzing the individual cells in these states, we found that in most cases, only one cell deviated significantly from preferred surface and volume, while the other cells had no residual stresses (i.e.\ their surface and volume deviations were below the respective cutoffs).  Interestingly, in almost all of these cases, the cell with residual stresses always had the shape of a so-called {\it truncated square trapezohedron} (TST)
(\fref{fig:g-residualStresses}b).  In our simulations, these TSTs always had a surface larger than $s_0$ and a volume smaller than $1$.   Moreover, they always had a surface-to-volume ratio of $s_i/v_i^{2/3}\approx 5.44364$.  

We tested whether $5.44364$ is the minimal surface-to-volume ratio that a TST can assume.  To this end, we derived analytical expressions for surface $S$ and volume $V$ of a TST, assuming 4-fold symmetry around the axis connecting the midpoints of the two squares and mirror-rotational symmetry perpendicular to this axis.  Our expressions for $S$ and $V$ contain three free parameters, one of which is a linear scaling parameter.  Numerical minimization of $s=S/V^{2/3}$ with respect to the other two parameters yielded indeed a minimum at $s_\mathrm{TST}=5.44363528(9)$.

This suggests the following picture for the vast majority of cases that are fluid while containing residual stresses.
Because $s_\mathrm{TST}$ is larger than the transition point $s_0^\ast$, for parameters $s_0$ in between both values, all cells can attain their preferred surfaces and volumes except for TST-shaped cells.  If a TST-shaped cell is contained in a configuration in this parameter regime, it will adjust surface area and volume to minimizes its own energy and will thus attain surface $s_i$ and volume $v_i$ such that $s_i/v_i^{2/3}=s_\mathrm{TST}$.  Its surface area $s_i$ will thereby be stretched above $s_0$ while its volume $v_i$ is compressed below $1$.  The actual value of $v_i$ depends on $k_V$.  For small $k_V$ the deviation $v_i-1$ will be larger than for large $k_V$.  

Note that theoretically, the fact that for $s_0<s_\mathrm{TST}$ TST-shaped cells attain a volume smaller than $1$ creates a slight shift in the transition point for these configurations.  This is because the non-TST cells now have to occupy a slightly larger volume.  Thus, they are also forced to attain a larger surface area. 
This induces an increase of the rigidity transition point by a very small offset. 

Also note that very rarely (3 times out of ca. 10,000), we have also encountered fluid configurations with residual stresses that were not explained by any of the reasons given in this section.  In these cases, many cells had surface and volume deviations above the cutoff in these cases. Due to numerical limitations, it was not possible to distinguish whether these cases represent real physics or just regions in the energy functional with very shallow gradients.

\subsection{The onset of residual stresses occurred in all cells at once.}
\label{sec:collectiveOnset}
\begin{figure}
  \includegraphics{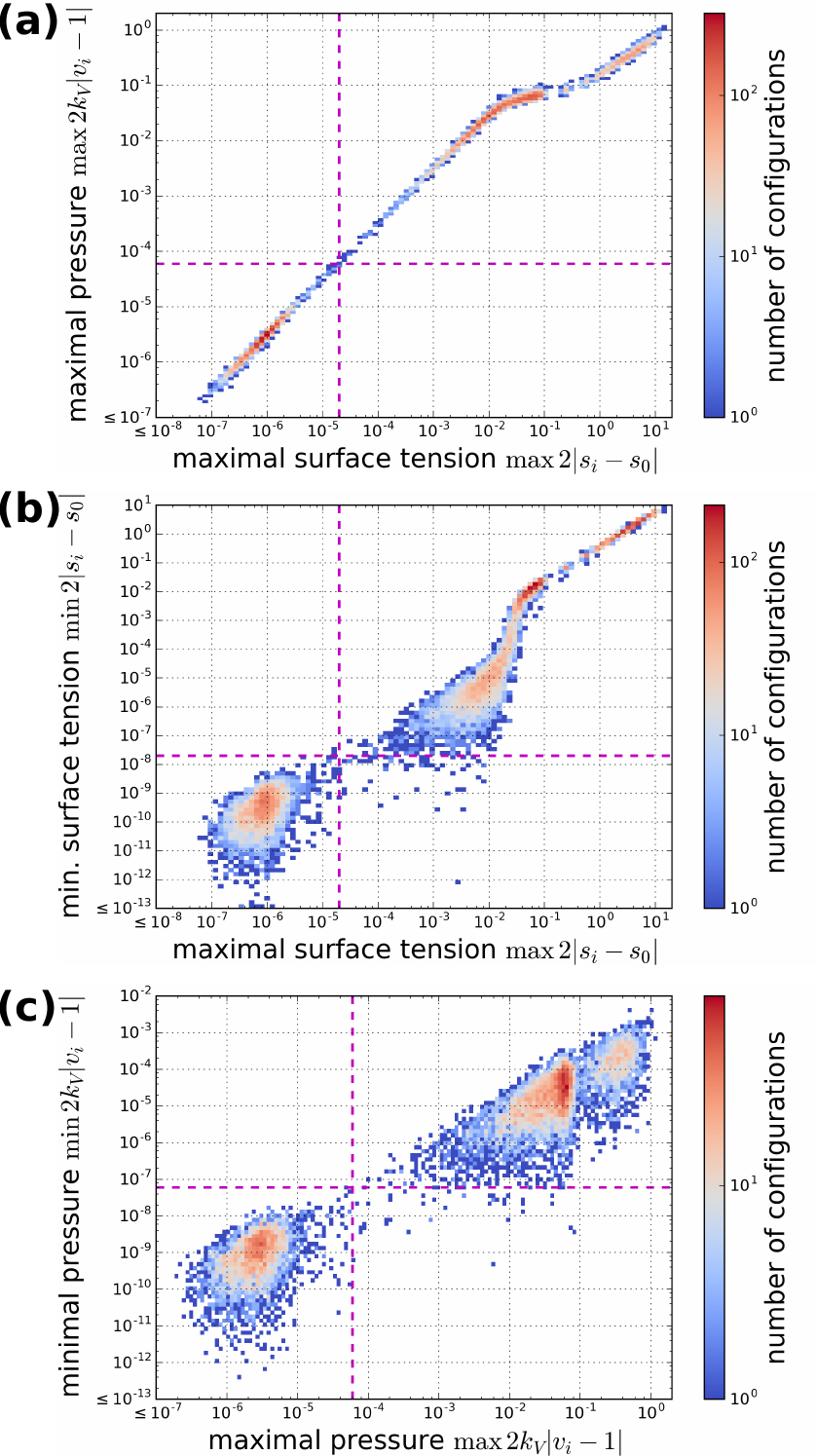}
  \caption{Collective onset of residual stresses, demonstrated using two-dimensional histograms of minimal and maximal surface tension and stresses in configurations with varying $s_0$ and $k_V=1$. 
  The horizontal and vertical magenta dashed lines indicate cutoffs defined in \ssref{sec:residualStressesCreateRigidity} and \ref{sec:collectiveOnset}.
  \label{fig:collectiveOnset}}
\end{figure}
Here we show that the rigidity transition was in our simulations a truly collective transition in the sense that the onset of residual stresses occurred in all cells at once.
To this end, using the two-dimensional histograms in \fref{fig:collectiveOnset}, we show that if one residual stress was non-vanishing, all were non-vanishing.  

First we show that whenever there was at least one non-vanishing cell surface tension, then there was a non-vanishing cell pressure, and vice versa.  To this end, we correlate maximum surface tension with maximum pressure (\fref{fig:collectiveOnset}a).
The horizontal and vertical magenta dashed lines represent again the respective cutoffs shown in \fsref{fig:g-residualStresses} and defined in \sref{sec:residualStressesCreateRigidity}.  Indeed, in this plot the upper-left and lower-right quadrants are completely empty, indicating that there are finite surface tensions if and only if there are finite pressures.

\begin{figure}
  \includegraphics{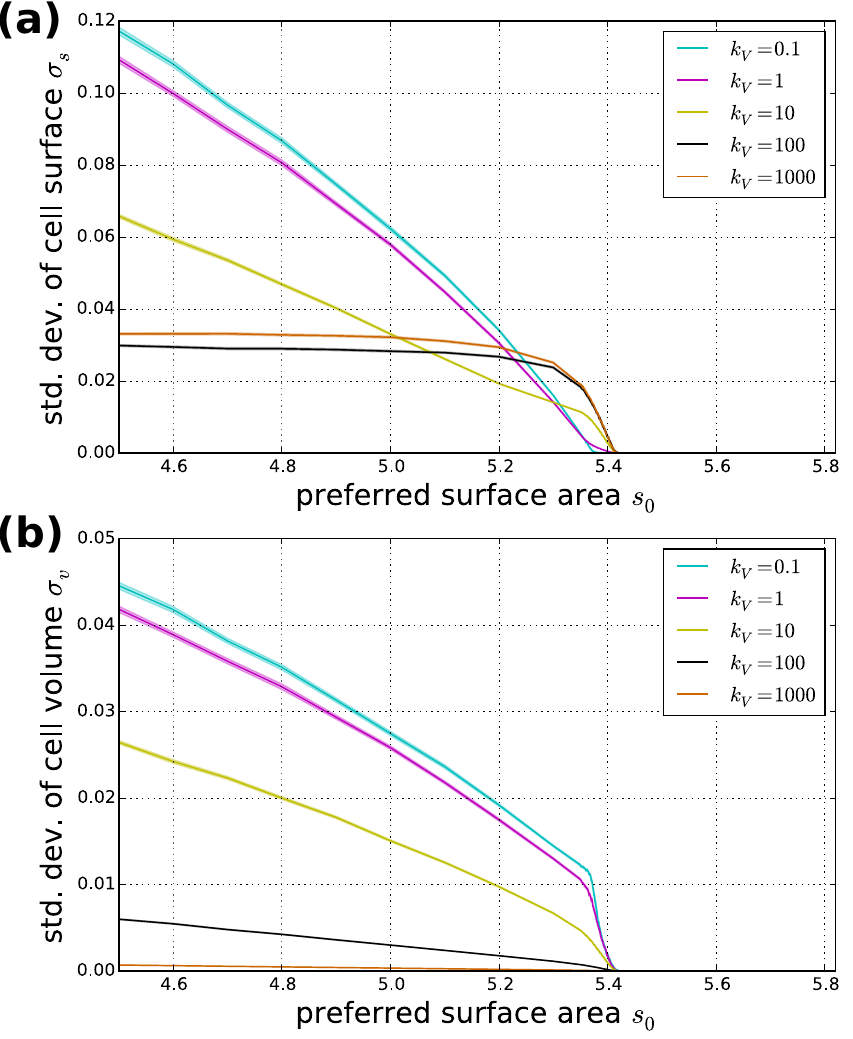}
  \caption{The surface and volume standard deviations are zero in the fluid and nonzero in the solid regime. 
  \label{fig:sigmas}}
\end{figure}
We next show that, whenever there was one non-vanishing cell surface tension, all cells had finite surface tensions.  We correlate maximal with minimal cell surface tension in \fref{fig:collectiveOnset}b.  The vertical dashed lines are again taken from \sref{sec:residualStressesCreateRigidity}.  We find that the fluid clusters can be clearly separated not only by a vertical, but also by a horizontal cutoff line, separating configurations where at least one cell has no surface tension from simulations where all cells have a finite surface tension.  The fact that only very few configurations are in the lower-right quadrant indicate that mostly whenever one cell deviated from its preferred surface, all cells did.
There are a few exceptions to these observations, appearing in the lower-right quadrant, which are typically due to the appearance of a TST-shaped cell (see previous section). 

Finally, we show that, whenever there was one finite cell pressure, all cell pressures were finite.  We correlate maximal with minimal cell pressure in \fref{fig:collectiveOnset}c.  We find again that the fluid clusters can be clearly separated by a horizontal cutoff line, separating configurations where at least one cell has no pressure from simulations where all cells have a finite pressure.  The fact that only few configurations appear in the lower-right quadrant indicates that except for a handful of exceptions,  whenever one cell deviated from its preferred volume, all cells did.  

These findings indicate that the rigidity transition in the 3D Voronoi model is indeed a truly collective one, where in the vast majority of cases, all cells simultaneously acquire both nonzero surface tension and nonzero pressure at the onset of rigidity.  This was true for all tested $k_V$ values between $0.1$ and $1000$.  However, it is in principle possible that these findings are due to finite-size effects.  We thus created plots similar to those in \fref{fig:collectiveOnset} for $N_c=4096$ cells.  Although we found that there were more configurations where only some of the cells had a finite surface tension that for $N_c=512$ ($\sim 1\%$ of all configurations), we found that in all these cases a TST-shaped cell with its minimal possible surface-to-volume ratio appeared. Thus, with the exception of the appearance of TSTs, even for $N_c=4096$ the onset of residual stresses is collective.

\subsection{Surface and volume standard deviations}
Here, we display the plots for the surface standard deviation $\sigma_s$ and volume standard deviation $\sigma_v$ (\fref{fig:sigmas}).  Note that both standard deviations vanish in the floppy regime, become nonzero at the transition point, and increase when going deeper into the solid regime.

\subsection{Scaling of the average cell surface tension in the solid vicinity of the transition point}
\label{sec:scalingSurfaceTension}
Here, we derive the scaling of the average surface tension $2(\langle s\rangle-s_0)$ as a function of the distance from the transition point $s_0^\ast$.
Because we study energy-minimized states and the energy can be written in terms of $\sigma_s$ and $\sigma_v$ (\eref{eq:energy-rewritten} with $\langle s\rangle=s_\mathrm{min}(\sigma_s,\sigma_v)$), we can have:
\begin{align}
  0 &= \frac{\partial e\Big(\langle s\rangle=s_\mathrm{min}(\sigma_s,\sigma_v),\sigma_s,\sigma_v\Big)}{\partial \sigma_s}\\
  0 &= \frac{\partial e\Big(\langle s\rangle=s_\mathrm{min}(\sigma_s,\sigma_v),\sigma_s,\sigma_v\Big)}{\partial \sigma_v}\text{.}  
\end{align}
Insertion of \esref{eq:energy-rewritten} and \seref{eq:sMin} yields:
\begin{align}
  \sigma_s &= a_s(\langle s\rangle-s_0)\\
  k_V\sigma_v &= a_v(\langle s\rangle-s_0)\text{.}  
\end{align}
Together with \eref{eq:sMin} and $\langle s\rangle=s_\mathrm{min}(\sigma_s,\sigma_v)$, we obtain indeed that:
\begin{equation}
  2(\langle s\rangle-s_0) = \frac{2\delta s_0}{1+a_s^2+a_v^2/k_V}\text{.}
\end{equation}

\bibliography{references}
\end{document}